\documentclass[iop]{emulateapj}
\usepackage{amsmath, amssymb}
\usepackage{amsmath, bm, mathtools, cancel, empheq, ulem, mathrsfs, natbib}
\usepackage{bm, graphicx}
\usepackage[counterclockwise]{rotating}
\usepackage[colorlinks]{hyperref}
\usepackage{multirow}
\usepackage[dvipsnames]{xcolor}
\usepackage[title]{appendix}
\hypersetup{
	colorlinks = true,
	linkcolor=blue,
	citecolor=blue
}
\usepackage{etoolbox}


\newcommand{\MYhref}[3][blue]{\href{#2}{\color{#1}{#3}}}

\newcommand{\ri}{r_{\rm{i}}}
\newcommand{\ro}{r_{\rm{o}}}

\newcommand{\cp}{c_{\rm{p}}}
\newcommand{\pderiv}[2]{\frac{\partial#1}{\partial#2}}
\newcommand{\sn}[2]{#1\times10^{#2}}

\newcommand{\five}{\ \ \ \ \ }
\newcommand{\rhobar}{\overline{\rho}}
\newcommand{\av}[1]{\langle#1\rangle}
\newcommand{\f}{\bm{\mathcal{F}}}
\newcommand{\I}{\mathscr{I}}
\newcommand{\raf}{{\rm{Ra}}}
\newcommand{\e}{\hat{\bm{e}}}
\newcommand{\newtext}{}

\slugcomment{The Astrophysical Journal, \MakeLowercase{000:00 (14pp), \today}}

\shorttitle{Revisiting the Sun's Strong Differential Rotation along Radial Lines}
\shortauthors{Matilsky et al.}

\begin{document}


\title{Revisiting the Sun's Strong Differential Rotation along Radial Lines}


\author{Loren I. Matilsky\altaffilmark{1}, Bradley W. Hindman, and Juri Toomre}
\affil{JILA \& Department of Astrophysical and Planetary Sciences, University of Colorado, Boulder, CO 80309-0440, USA}

\altaffiltext{1}{loren.matilsky@colorado.edu}


\begin{abstract}
Current state-of-the-art models of the solar convection zone consist of solutions to the Navier-Stokes equations in rotating, 3D spherical shells. Such models are highly sensitive to the choice of boundary conditions. Here, we present two suites of simulations differing only in their outer thermal boundary condition, which is either one of fixed-entropy or fixed-entropy-gradient. We find that the resulting differential rotation is markedly different between the two sets. The fixed-entropy-gradient simulations have strong differential rotation contrast and isocontours tilted along radial lines (in good agreement with the Sun's interior rotation revealed by helioseismology), whereas the fixed-entropy simulations have weaker contrast and contours tilted in the opposite sense. We examine in detail the force balances in our models and find that the poleward transport of heat by Busse columns drives a thermal wind responsible for the different rotation profiles. We conclude that the Sun's strong differential rotation along radial lines may result from the solar emissivity being invariant with latitude (which is similar to the fixed-entropy-gradient condition in our models) and the poleward transport of heat by Busse columns. In future work on convection in the solar context, we strongly advise modelers to use a fixed-gradient outer boundary condition.
\end{abstract}




\section{Introduction}
\begin{figure}
	\includegraphics[width=3.4375in]{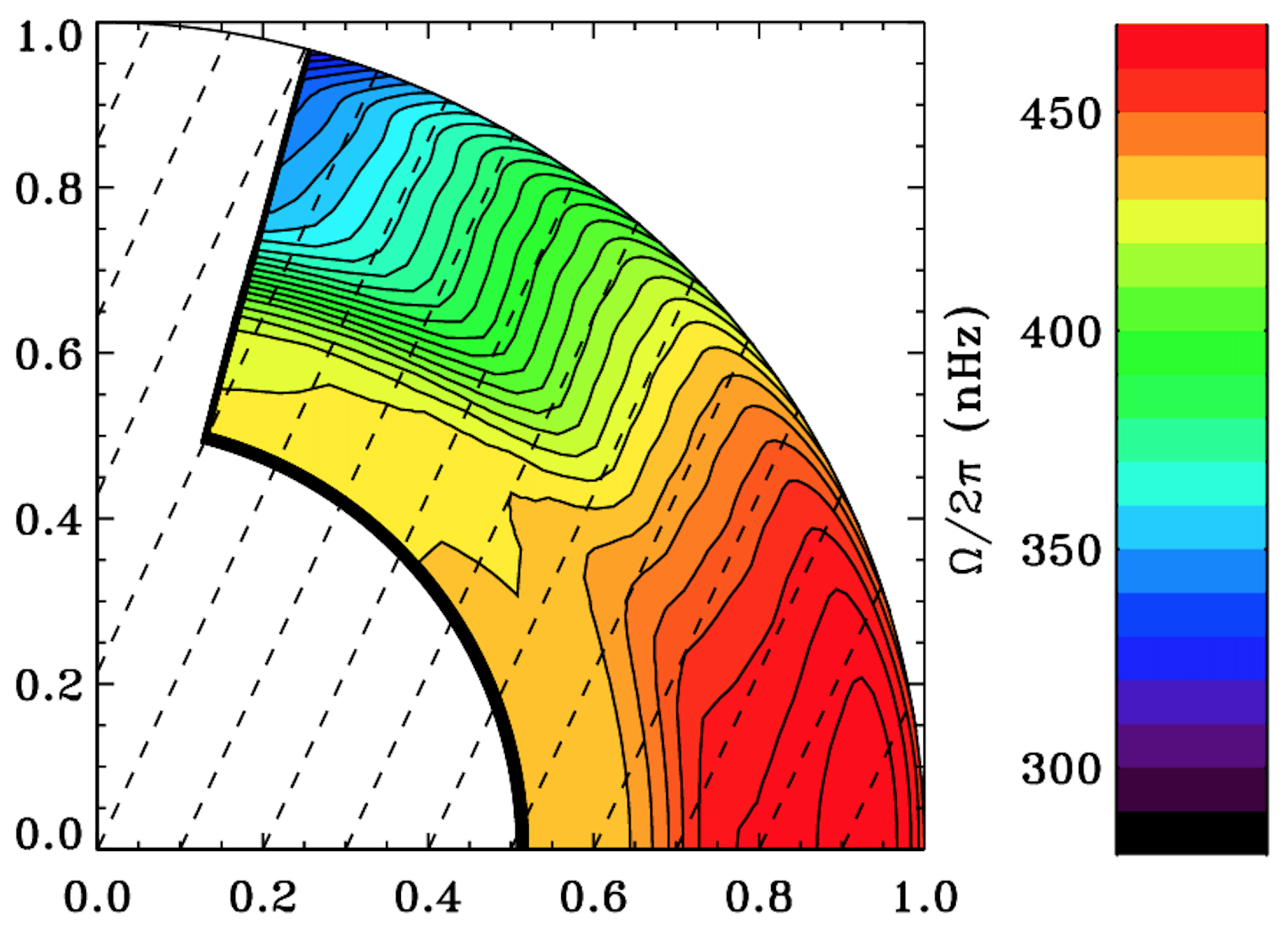}
	\caption{Contour map in the upper meridional plane of the internal rotation profile of the Sun, in and below the CZ, averaged in longitude and time from 1995--2009. The rotation rate has been obtained using a regularized least-squares (RLS) inversion, which is sensitive only to the equatorially symmetric part of the rotation. The dashed lines are at a 25$^\circ$ angle to the rotation axis and align with the isorotation contours at mid-latitudes. Image credit: \citet{Howe05}, extended to include GONG data until 2009.}
	\label{fig:gong}
\end{figure}

Helioseismology has revealed in detail the internal rotation profile of the solar convection zone (CZ; e.g., \citealt{Thompson03}; \citealt{Howe05}), as shown in Figure \ref{fig:gong}. The most notable properties of the rotation rate are that the equator rotates significantly faster than the high-latitude regions and that the isorotation contours are tilted significantly with respect to the rotation axis, falling largely along radial lines. Furthermore, there are two shear layers at the top and bottom of the CZ: at the top, the contours bend toward the equator in a region known as the \textit{near-surface shear layer (NSSL)}, and at the bottom, the differential rotation in the CZ transitions to solid-body rotation, over a narrow boundary layer called the \textit{tachocline}. Prior to helioseismic probing, most theoreticians had assumed that the differential rotation that is observed directly at the surface  would imprint into the interior along isosurfaces parallel to the rotation axis, hence satisfying the Taylor-Proudman theorem. The helioseismic observations have clearly demonstrated that this theoretical supposition was wrong.

For the last several decades, global, 3D supercomputer simulations of hydrodynamic convection in rotating spherical shells have succeeded in achieving rotation profiles that are fast at the equator and slow at the poles. However, simulations generally have a weaker overall differential rotation contrast than that of the Sun. If the contrast is defined to be the difference in rotation rate between the equator and $60^\circ$ latitude, expressed as a percentage of the ``frame" rotation rate, then for the Sun, this magnitude is $\sim$$20\%$. Most simulations, on the other hand, have rotation-contrast magnitudes of $\sim$$10\%$, although there are some notable exceptions (e.g., \citealt{Brun02}; \citealt{Brown10}; \citealt{Matilsky20}). To date, however, there is no systematic physical explanation for why these particular simulations have high rotation contrast.

Simulations have also struggled to achieve rotation contours in the bulk of the CZ that are significantly tilted from the axis, as seen in Figure \ref{fig:gong}. With some exceptions (e.g., \citealt{Elliot00};  \citealt{Miesch06}), the simulations generally have cylindrically-aligned contours. In the case of \citet{Miesch06}, the contours were tilted systematically by imposing a modest latitudinal entropy gradient at the base of the CZ. By modifying the magnitude of this entropy gradient, \citet{Miesch06} controlled the magnitude of the contour tilt. Imposing this boundary condition was inspired by thermal-wind balance in the tachocline imprinting up into the CZ.

\newtext{There is also good evidence for thermal-wind balance operating in the deep CZ from analytical models. \citet{Balbus09a, Balbus09b} assumed a simple functional relationship between the entropy and angular velocity and solved explicitly for the combined isorotational and isentropic contours as characteristics of the thermal-wind equation. This yielded tilted contours in good agreement with helioseismology for the bulk of the CZ (away from the boundary layers of shear at the top and bottom). Such analytical work---in addition to lending strong support for the presence of a thermal wind in the solar interior---points to the need to better understand the dynamical coupling between entropy and angular velocity in global numerical simulations, something we address in the present paper.}

\newtext{In this work, we explore the role that the outer thermal boundary condition plays, in conjunction with an interior thermal wind, in modifying the resulting differential rotation.} We consider the two most commonly used options: fixed-entropy (FE), in which the entropy at the outer boundary is fixed to a constant value, and fixed-flux (FF), in which the radial entropy gradient at the outer boundary is fixed, therefore implying an outward conductive flux that is independent of latitude. \newtext{See \citet{Hurle66} and \citet{Edwards90} for descriptions of these boundary conditions in the context of linear theory, and \citet{Anders20} for a detailed analysis of the boundary conditions in nonlinear simulations of Rayleigh-B\'enard convection.} The FF condition is more appropriate for the Sun, since the radiant flux from the solar photosphere does not appear to vary substantially with latitude. \citet{Rast08} analyzed full-disk images from the Precision Solar Photometric Telescope at the Mauna Loa Solar Observatory and calculated non-magnetic contributions to the solar photospheric intensity. In both continuum and Ca II K intensity distributions, only a $\sim$0.1--0.2\% variation was observed, corresponding to a solar pole that is at most $\sim$2.5 K warmer (in terms of effective temperature) than the equator. \newtext{The deduced near-spherical-symmetry of the solar emissive flux is a significant observation, since a thermal wind strong enough to drive the observed differential rotation would require greater differences in the interior temperature between equator and pole. We return to this point in our concluding remarks.}

We do not address the dynamics of the near-surface nor tachocline shear layers in this work. In particular, our models have an impenetrable lower boundary that does not allow for the convective overshoot of downflow plumes into the stable region that may play a role in the origin of the tachocline. The dynamical maintenance of the NSSL is still an open question, as discussed in \citet{Hotta15} and \citet{Matilsky19}, and models tend to only display signs of near-surface shear if they have high density contrast \newtext{($>\sim$100)} across the CZ. To avoid high computational cost, the models in this work \newtext{have a smaller density contrast of $\sim$20}. 

Solar-like differential rotation (fast equator and slow poles) in spherical-shell convection is thought to be due to the outward transport of angular momentum by Busse columns (also called ``Taylor columns" and ``banana cells"; e.g., \citealt{Busse02}; \citealt{Jones11}). In this work, we show that Busse columns also transport heat poleward and equivalently drive a solar-like differential rotation through a thermal wind. We find that  in our FF cases, the thermal wind drives stronger differential rotation magnitudes and achieves more significant tilt in the rotation contours than the corresponding FE cases. \citet{Elliot00} noted this effect for one simulation, but did not explore the underlying mechanism. 

In Section \ref{sec:num}, we describe the parameter space explored by our simulation suite, as well as the mathematical details of the FF and FE boundary conditions. In Section \ref{sec:res}, we describe the basic results of our experiment, focusing on the achieved differential rotation. In Section \ref{sec:tw}, we quantify the force balance achieved in our models, which, for the radial and latitudinal directions, consists of a thermal wind in spherical geometry. In Section \ref{sec:en}, we examine the latitudinal transport of energy by Busse columns that is responsible for the thermal wind. In Section \ref{sec:bc}, we discuss how the effects of the thermal wind are modified by the outer thermal boundary condition. In Section \ref{sec:disc}, we discuss our simulation results in the context of the Sun. 

\section{Numerical Experiment}\label{sec:num}
We consider time-dependent, 3D simulations of a rotating, stratified spherical shell of fluid representative of the solar CZ. We use the open-source code Rayleigh 0.9.1 \citep{Featherstone16a, Matsui16, Featherstone18}, which solves the equations of hydrodynamics in spherical geometry. Our domain is a spherical shell with inner radius $\ri$ and outer radius $\ro$. We describe this domain using the spherical coordinates ($r,\theta,\phi$) and corresponding unit vectors ($\hat{\bm{e}}_r$, $\hat{\bm{e}}_\theta$, $\hat{\bm{e}}_\phi$). \newtext{Here, $r$ is the spherical radius, $\theta$ the colatitude and $\phi$ the azimuthal coordinate.} When convenient, we also use cylindrical coordinates $(\lambda, \phi, z) = (r\sin\theta,\phi,r\cos\theta)$ and unit vectors ($\hat{\bm{e}}_\lambda$, $\hat{\bm{e}}_\phi$, $\hat{\bm{e}}_z$), where $\lambda$ is the cylindrical radius and $z$ the axial coordinate perpendicular to the equatorial plane.

The thermodynamic reference state is chosen to be temporally steady and spherically symmetric with adiabatic stratification (see \citealt{Jones11} for a complete description). The density varies by a factor of  $\sim$20 (three density scale heights) across the layer. We denote the pressure, density, temperature, and entropy by $P$, $\rho$, $T$, and $S$, respectively, using overbars to indicate the fixed reference state and the absence of overbars to indicate deviations from the reference state. 

Rayleigh uses an anelastic approximation (e.g., \citealt{Gough69}; \citealt{Gilman81}; \citealt{Jones11}), which removes sound waves from the system, making the maximum allowable timestep much larger since it is limited by the flow velocity rather than the sound speed. In a frame rotating with angular velocity $\bm{\Omega}_0 = \Omega_0\e_z$, the fluid equations are then given by (e.g., \citealt{Featherstone16a})
\begin{align}
\nabla\cdot(\overline{\rho}\bm{v}) &=  0,
\label{eq:cont}
\end{align}
\begin{align}
\overline{\rho}\Bigg{[}\frac{\partial\bm{v}}{\partial t} + (\bm{v}\cdot\nabla)\bm{v}\Bigg{]} = &-2\overline{\rho}\bm{\Omega}_0\times\bm{v} \nonumber\\
&-\overline{\rho}\nabla \Bigg{(}\frac{P}{\overline{\rho}}\Bigg{)} -\frac{\overline{\rho} S}{\cp}\bm{g}
+ \nabla\cdot \bm{D},
\label{eq:mom}
\end{align}
and
\begin{align}
\overline{\rho}\overline{T}\Bigg{[}\frac{\partial S}{\partial t} + \bm{v}\cdot\nabla S\Bigg{]} =\ &\nabla\cdot\big{[}\kappa\overline{\rho}\overline{T}\nabla S\big{]} + Q + \bm{D}:\nabla\bm{v}.
\label{eq:en}
\end{align}
Here, $\bm{v} = (v_r, v_\theta, v_\phi)$ is the fluid velocity in the rotating frame, $\cp$ is the constant-pressure specific heat, $\bm{g}$ is the gravitational acceleration due to a solar mass $M_\odot$ located at the center of the spherical shell, $\bm{D} = \rhobar\nu[\nabla\bm{v} + \nabla\bm{v}^T - (2/3)(\nabla\cdot\bm{v})\bm{I}]$ is the Newtonian viscous-stress tensor, $\bm{I}$ is the identity tensor, $\nu$ is the kinematic viscosity, and $\kappa$ is the thermal diffusivity. Because it is not computationally possible to resolve convection in the solar regime down to the turbulent microscale, $\nu$ and $\kappa$ must be regarded as ``eddy" diffusivities, which for simplicity  we choose to be spatially constant and equal, such that the Prandtl number is unity. The internal heating function, which physically represents heating due to radiation, is chosen to have the fixed radial profile $Q(r)=\alpha[\overline{P}(r) - \overline{P}(\ro)]$, with the normalization constant $\alpha$ chosen such that a solar luminosity $L_\odot$ is forced through the domain.

\newtext{The heating function $Q$ is designed such that most of the energy is deposited in the bottom $\sim$1/3 of the domain (see \citealt{Featherstone16a} for the radial profile of the heat flux associated with $Q$). In the Sun, the internal heating arises from high-opacity metals in the CZ absorbing the radiation streaming in from the interior. This internal heating is quite distributed, and the heat flux associated with our imposed $Q$ agrees well with the radiative flux inferred from standard solar models (for example, model S in \citealt{Dalsgaard96}). The distributed nature of the heating makes the convection problem in the solar CZ slightly different from a standard Rayleigh-B\'enard setup, in which the fluid layer is heated from below by imposing a conductive flux through the lower boundary.}

The equation of state for the system is that of a perfect gas subject to small thermodynamic perturbations about the reference state:
 \begin{align}
\frac{\rho}{\overline{\rho}} = \frac{P}{\overline{P}} - \frac{T}{\overline{T}} = \frac{P}{\gamma\overline{P}}- \frac{S}{\cp},\label{eq:eos}
\end{align}
where $\gamma=5/3$ is the ratio of specific heats.

We adopt stress-free and impenetrable boundary conditions to conserve angular momentum and mass:
\begin{align}
v_r = \pderiv{}{r}\bigg{(}\frac{v_\theta}{r}\bigg{)}  = \pderiv{}{r}\bigg{(}\frac{v_\phi}{r}\bigg{)} = 0\ \text{at}\ r=\ri\ \text{and}\ \ro\label{eq:vbc}.
\end{align}

In all cases, the inner thermal boundary condition allows no flux of energy into the system through the lower boundary:
\begin{align}
 \pderiv{S}{r} = 0\ \text{at}\ r=\ri. 
 \label{eq:sbc_inner}
 \end{align}
 Input parameters common to all the simulations explored here are shown in Table \ref{tab:input}.
 \begin{table}
 	\caption{Common input-parameter values for all simulations}\label{tab:input}
 	\centering
 	\begin{tabular}{r  l}
 		\hline\hline
 		$\ri$       & 5.00 $\times10^{10}$ cm = $0.719\ R_\odot$\\
 		$\ro$  & 6.59 $\times10^{10}$ cm = $0.947\ R_\odot$\\
 		$\cp$ & 3.50 $\times10^8$ erg K$^{-1}$ g$^{-1}$\\
 		$\gamma$  & 1.67 \\
 		$\overline{\rho}_{\rm{i}}$ & 0.181 g cm$^{-3}$ \\
 		$L_\odot$ & 3.85 $\times10^{33}$ erg s$^{-1}$\\
 		$M_\odot$ & 1.99 $\times10^{33}$ g\\
 		$R_\odot$ & 6.96 $\times10^{10}$ cm\\
 		Pr $\equiv$ $\nu/\kappa$ & 1.00 \\
 		\hline
 	\end{tabular}
 \end{table}

\subsection{Outer thermal boundary condition}
 The main purpose of this work is to characterize the influence of the outer thermal boundary condition on the behavior of the resulting differential rotation. We consider models with different background rotation rates $\Omega_0$ and diffusion values ($\nu=\kappa$) and for each model analyze two sub-cases:
 \begin{align}
 S = 0\ \text{at}\ r=\ro \five \text{(fixed entropy, or FE)}\label{eq:bcfixeds}
 \end{align}
 and
  \begin{align}
 \pderiv{S}{r} = -\frac{L_\odot}{4\pi\ro^2 \rhobar\overline{T}\kappa}\ \text{at}\ r=\ro \five \text{(fixed flux, or FF)}\label{eq:bcfixedflux}.
 \end{align}

The solar luminosity that is injected into the system via internal heating is ultimately carried out through the outer surface via thermal conduction, which in our models arises from entropy gradients (see Equation \eqref{eq:en}). For the fixed-entropy condition \eqref{eq:bcfixeds}, the interior is initially heated (leading to $S>0$ in the lower parts of the CZ) while the entropy at the outer surface is ``pinned" to zero. This naturally establishes a thermal boundary layer (sharp entropy gradients $\partial S/\partial r < 0$) just below the outer surface. The steepness of the gradient (i.e., the strength of the outward conductive loss of energy) is allowed to vary with latitude. 

For the fixed-flux condition \eqref{eq:bcfixedflux}, the outer thermal boundary layer is present from the beginning of the simulation. The steepness of the entropy gradient (and thus the energy loss) at the outer surface is independent of latitude by construction, and is forced to have exactly the value needed to carry out a solar luminosity. The fixed-flux condition is thus more ``solar-like," since in the Sun there is no observed latitudinal dependence of the emergent intensity, which is equal to the energy lost via radiative cooling at the photosphere. 

For both the FE and FF cases, the thermal conductive boundary layer stands in contrast to the real solar photosphere, in which radiative cooling removes the heat from a very thin ($\sim$100 km) outer layer. The cooling drives very small temporal and spatial scales of motion compared to the deep interior (such as granulation and supergranulation), making its direct inclusion in global models problematic. Researchers have sought to address this difficulty in various ways. \citet{Nelson18} implemented stochastic driving of convection by near-surface plumes designed to mimic the effects of supergranulation, finding that that the flow structures and transport properties were significantly altered in the deep CZ. \citet{Hotta19} simulated the whole CZ with no rotation or magnetic field, coupling a global spherical shell that captured large-scale flows in the deep interior to a Cartesian box that solved the equations of radiative transfer in the photosphere. They found that the near-surface motions had a weak influence on the deep interior. Regardless of its relevance to interior flow structures, correctly capturing the small-scale near-surface flows in global models is currently prohibitively expensive computationally. In order to explore a wider range of parameter space, we thus only consider the FE and FF boundary conditions here. 

 \section{Simulation Results}\label{sec:res}
  We label simulations with a prefix that signifies the outer boundary condition (``FE" for Equation \eqref{eq:bcfixeds} and ``FF" for Equation \eqref{eq:bcfixedflux}), followed by the value of the diffusion constant $\nu=\kappa$ (in units of $10^{12}$ cm$^2$ s$^{-1}$), followed by the value of the rotation rate (in units of the sidereal Carrington value for the Sun, $\Omega_\odot\equiv 2.87\times10^{-6}$ rad s$^{-1}$, or $\Omega_\odot/2\pi \equiv 456\ \rm{nHz}$). For example, ``case FE4-3" refers to a simulation with an FE outer boundary, for which $\nu=\kappa=4\times10^{12}$ cm$^2$ s$^{-1}$ throughout the domain, and $\Omega_0=3\Omega_\odot$.
  
 Table \ref{tab:output} \newtext{(in the Appendix)} contains the values of the non-dimensional parameters, as well as the grid resolution, for each of the 18 simulations considered in this work. \newtext{Table \ref{tab:output}  has four groupings according to FE or FF  at two different rotation rates.} Following the notation of \citet{Featherstone16b}, we parameterize the strength of the imposed driving in each simulation through a bulk ``flux Rayleigh number" (imposed a priori),
 \begin{align}
 \raf \equiv \frac{\tilde{g}\tilde{F}H^4}{\cp\tilde{\rho}\tilde{T}\nu^3}
 \label{def:raf}
 \end{align}
 \newtext{and the level of turbulence through bulk Reynolds or P\'eclet numbers (calculated a posteriori),}
 \begin{align}
 \text{Re} = \frac{\text{Pe} }{\text{Pr}} =  \frac{\tilde{v}^\prime H}{\nu}.
 \end{align}
 \newtext{Since the Prandtl number for all models is unity, the P\'eclet number Pe $=\tilde{v}^\prime H\kappa^{-1}$ equals the Reynolds number.}
 
 Similarly, we parameterize the influence of rotation through an Ekman number (imposed a priori),
  \begin{align}
 {\rm{Ek}} \equiv \frac{\nu}{2\Omega_0 H^2}
 \label{def:ek}
 \end{align}
  and a bulk Rossby number (calculated a posteriori),
 \begin{align}
 {\rm{Ro}} \equiv \frac{\tilde{v}^\prime}{2\Omega_0 H}.
 \label{def:ro}
 \end{align}

 In the preceding equations, the length scale $H$ is taken to be the shell depth $\ro - \ri$, the tildes refer to volume averages over the full spherical shell, and $F$ refers to the energy flux associated with conduction and convection in equilibrium (see \citealt{Featherstone16b}). The typical convective velocity amplitude $\tilde{v}^\prime$ refers to the rms of the velocity with the longitudinally averaged part subtracted, the mean being taken over time and over the full volume of the shell. Throughout this work, temporal averages are taken during the latter portion of run time for which there is statistical equilibrium---generally $\sim$3/4 of the total run time listed in Table \ref{tab:output}.
 
 \newtext{Before discussing our results in detail, we note that all our models have fairly high levels of thermal and viscous diffusion. Furthermore, all our models rotate at either two or three times the solar Carrington rate. These choices, which stand in contrast to the physics of the solar interior, ensure that our models have low enough Rossby numbers to avoid antisolar differential rotation (fast poles, slow equator). All global spherical-shell convection codes  produce high velocities at large scales in the solar context when sufficiently turbulent.  The influence of rotation on the large scales is therefore weak, which causes less coherence in the outward angular momentum transport by the convection, and ultimately less angular momentum in the outer layers than the inner layers (i.e., an antisolar differential rotation). The overall problem---that increasing the turbulence in simulations leads to antisolar states---is now called the ``convective conundrum" \citep{OMara16}. The antisolar states can be avoided by raising the rotation rate, raising the diffusions, or lowering the luminosity. We choose a combination of the former two, which requires that our models are only moderately turbulent. Nonetheless, the viscous force and heat flux are small to leading order in the primary dynamical balances.}
 
 \newtext{Returning to our simulation results,} we quantify the magnitude of the overall differential rotation contrast as the difference in the outer-surface rotation rate between the equator and 60$^\circ$-latitude, normalized by the frame rotation rate:
\begin{align}
\frac{\Delta\Omega}{\Omega_0}\equiv \frac{\Omega(\ro, \pi/2) - \Omega(\ro, \pi/6)}{\Omega_0},
\label{def:drf}
\end{align}
 where 
\begin{align}
\Omega(r,\theta)\equiv \Omega_0 + \frac{\av{v_\phi}}{r\sin{\theta}}
\label{def:om}
\end{align}
is the rotation rate of the fluid as a function of $r$ and $\theta$ and the angular brackets denote a combined temporal and longitudinal average. \newtext{From Table \ref{tab:output}, the FF cases have differential rotation contrasts $\Delta\Omega/\Omega_0$ that are significantly greater---on average, by $\sim$$40\%$---than those of the FE cases. For comparison, the solar value of the rotation contrast is substantially higher than in any of our models: $\Delta\Omega_\odot/\Omega_\odot = 0.20$ (see \citealt{Howe00}, Figure 1). For the solar estimate, we have taken $\Omega_\odot$ to be the sidereal Carrington rate and $\ro$ to lie just below the near-surface shear layer.}

\begin{figure}
	\includegraphics{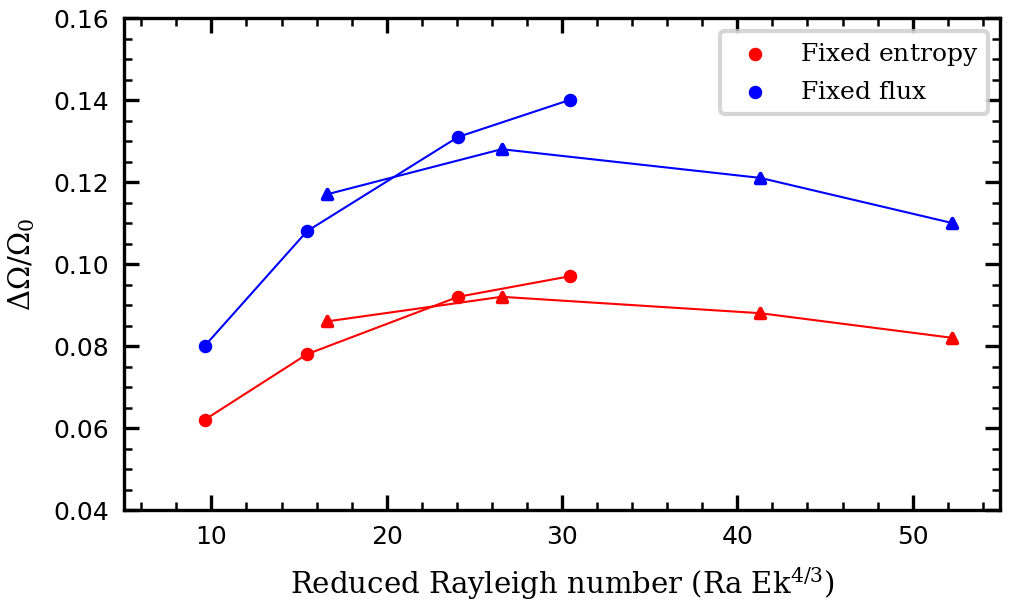}
	\caption{Rotation contrast achieved by the suite of simulations plotted as  a function of the reduced Rayleigh number $\mathcal{R} \equiv\rm{Ra\ Ek^{4/3}}$. \newtext{Circles mark the $\Omega_0=3\Omega_\odot$ cases and triangles the $\Omega=2\Omega_\odot$ cases.} }
	\label{fig:contrast_vs_Rayleigh}
\end{figure}

Figure \ref{fig:contrast_vs_Rayleigh} shows how the differential rotation fraction scales with the reduced Rayleigh number, which accounts for   the increase to the critical Rayleigh number for convective onset caused by rotation \citep{Chandrasekhar61}. The reduced Rayleigh number thus serves as a better parameterization of the supercriticality of the system than simply the Rayleigh number. \newtext{From Figure \ref{fig:contrast_vs_Rayleigh}, each type of boundary condition yields a similar scaling with the reduced Rayleigh number $\mathcal{R}$. For the $\Omega_0=3\Omega_\odot$ cases (circles in Figure \ref{fig:contrast_vs_Rayleigh}), the rotation contrast increases monotonically (but with decreasing slope) with increasing $\mathcal{R}$, so that the curves connected by circles in Figure \ref{fig:contrast_vs_Rayleigh} are concave-down. For the $\Omega_0=2\Omega_\odot$ cases (triangles in Figure \ref{fig:contrast_vs_Rayleigh}), the curves ``overturn" so that a peak value of the rotation contrast (at around $\mathcal{R}\sim27$) is achieved. This behavior (concisely described in \citealt{Gastine13}) is a symptom of the convective conundrum; as models grow more turbulent, the rotation contrast increases at first, but then decreases and eventually becomes negative (i.e., antisolar).}

\begin{figure}
	\includegraphics{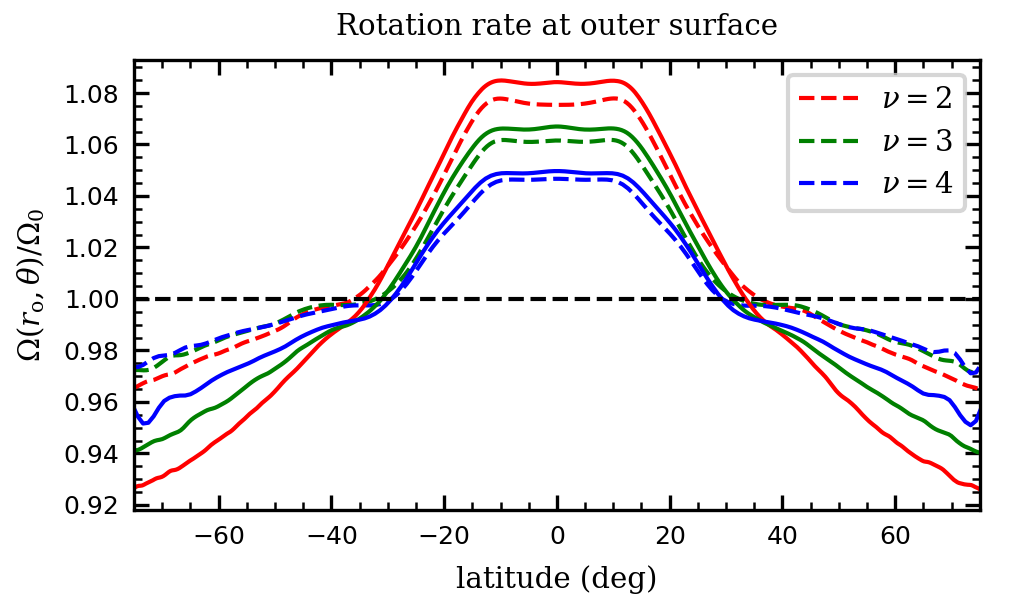}
	\caption{Temporally and longitudinally averaged rotation rate (\newtext{normalized by the frame rate $\Omega_0$}) at the outer surface for three of the cases rotating at $\Omega_0 = 3\Omega_\odot$, plotted versus latitude. Dashed lines correspond to the FE cases and solid lines to the FF cases. In the legend, $\nu=\kappa$ is given in units of $10^{12}\ \rm{cm}^2\ \rm{s}^{-1}$.}
	\label{fig:diffrot_vs_theta}
\end{figure}

To illustrate exactly where the ``extra" rotation contrast in the FF cases is located, we plot the rotation rate at the outer surface for three of the cases rotating at $3\Omega_\odot$ in Figure \ref{fig:diffrot_vs_theta}. Most of the additional contrast occurs at high latitudes, where the polar regions in the FF cases rotate significantly more slowly than in their FE counterparts. Additionally, the equator in the FF cases rotates slightly faster than in the FE cases. For all simulation pairs, the difference in contrast between the FE case and the FF case is greater the smaller the value of the diffusion (or the higher the level of turbulence). 

\begin{figure*}
	\includegraphics{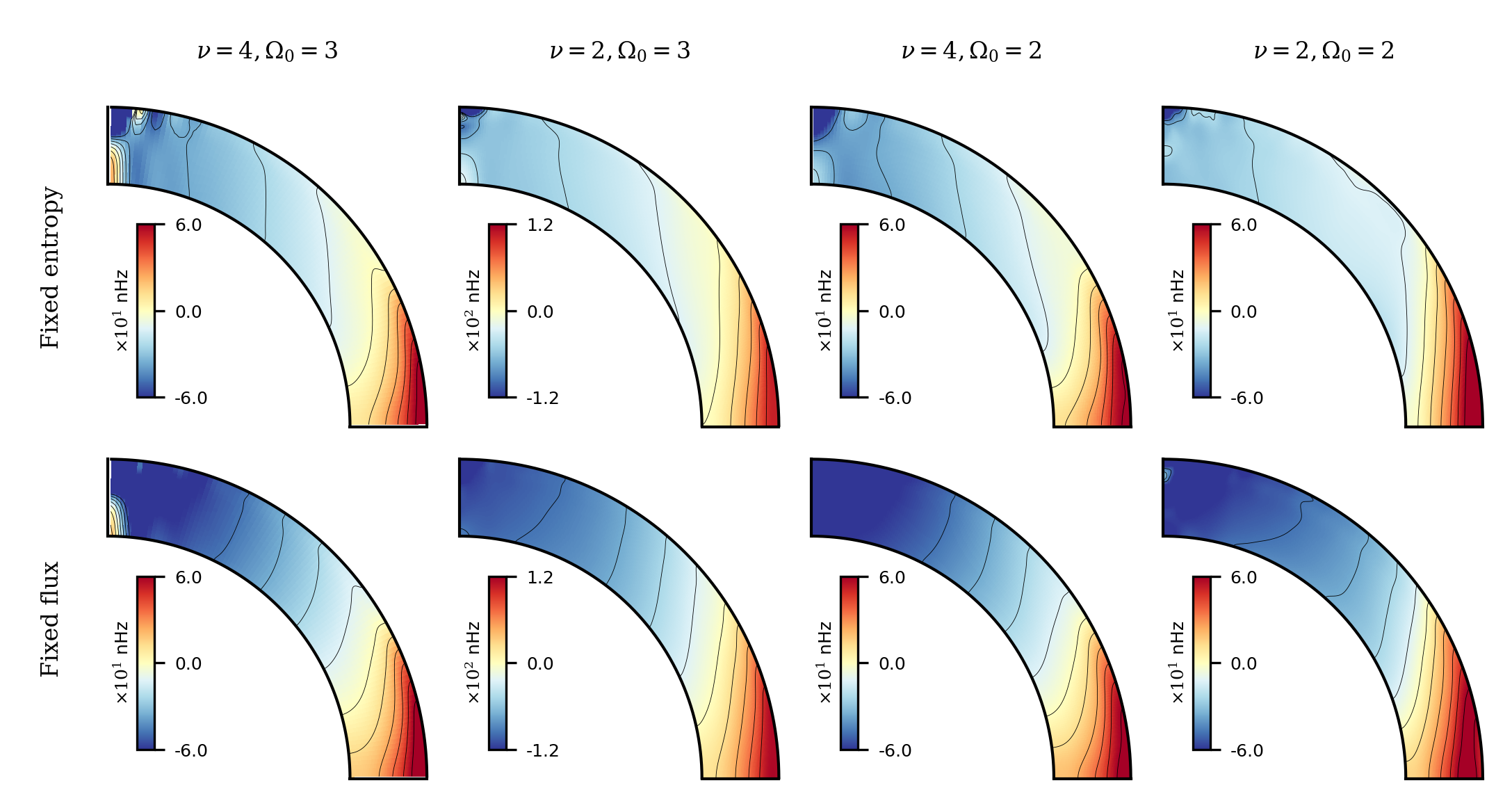}
	\caption{Temporally and longitudinally averaged rotation rate in the meridional plane for some of the simulations in the suite, with the two hemispheres averaged assuming even symmetry about the equator. In the titles at the top, $\nu=\kappa$ is given of units of $10^{12}$ cm$^2$ s$^{-1}$, and $\Omega_0$ in units of $\Omega_\odot$.}
	\label{fig:diffrot_halfplane}
\end{figure*}
Figure \ref{fig:diffrot_halfplane} shows contour plots of rotation rate in the upper meridional plane for some of the simulation suite. Clearly there is a striking difference between the tilts of the rotation contours in the FE and FF simulations. 

\begin{figure}
	\includegraphics[width=3.4375in]{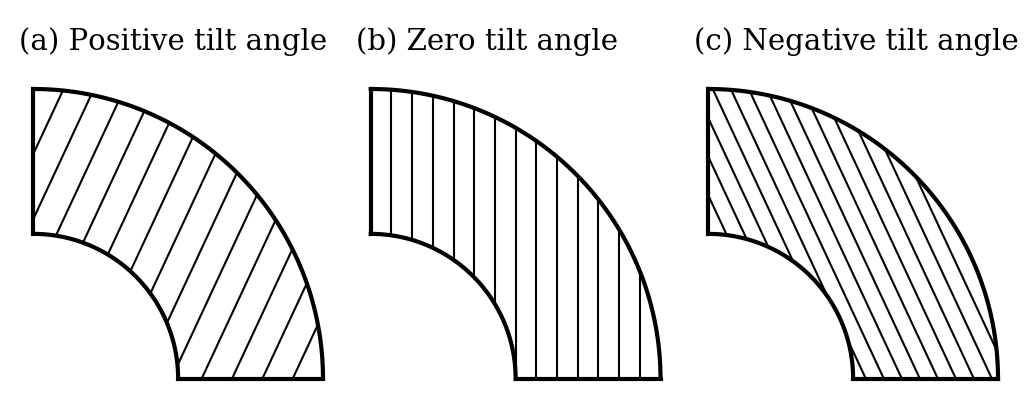}
	\caption{Schematic for our definition of contour tilts, showing (\textit{a}) positive contour tilt (all contours tilted at a constant $+25^\circ$), (\textit{b}) zero contour tilt, and  (\textit{c}) negative contour tilt (all contours tilted at a constant $-25^\circ$).}
	\label{fig:schematic_tilt}
\end{figure}

In this paper, we define all rotation-contour \textit{tilt angles} (or simply \textit{tilts}) with respect to the rotation axis, with zero tilt corresponding to alignment of the contour with the rotation axis. We use the sign convention for tilt angle illustrated in Figure \ref{fig:schematic_tilt}. Under this convention, the solar rotation contours have positive tilts at all latitudes in the bulk of the CZ (above the tachocline and below the NSSL, as shown in Figure \ref{fig:gong}). We thus define the tilt angle of a rotation contour at a given point in the meridional plane as
\begin{align}\label{eq:tilt}
{\rm{tilt}} \equiv -\tan^{-1}\bigg{[}\frac{\partial\Omega/\partial z}{\partial\Omega/\partial \lambda}\bigg{]},
\end{align}
which is consistent with the sign convention shown in Figure \ref{fig:schematic_tilt} for solar-like differential rotation, in which the contours further from the rotation axis correspond to a higher rotation rate.  

Describing the solar rotation contours as ``tilted along radial lines," as is often done, is technically misleading. Radial tilt implies a specific dependence of the contour tilt angle with latitude, namely, contours that fan radially outward from the center of the Sun. In Figure \ref{fig:gong}, by contrast, the bulk-CZ tilts are roughly constant at $\sim$25$^\circ$ for mid-latitudes, are smaller than $\sim$25$^\circ$ for low latitudes (where radially-aligned tilts would be greater), and are greater than $\sim$25$^\circ$ for high latitudes (where radially-aligned tilts would be smaller). To avoid confusion, we will henceforth not use the term ``radial tilt" and instead describe the rotation-contour tilt (in the Sun and in our simulations) simply as ``positive" or ``negative," using the sign convention illustrated in Figure \ref{fig:schematic_tilt}.

\begin{figure}
	\includegraphics{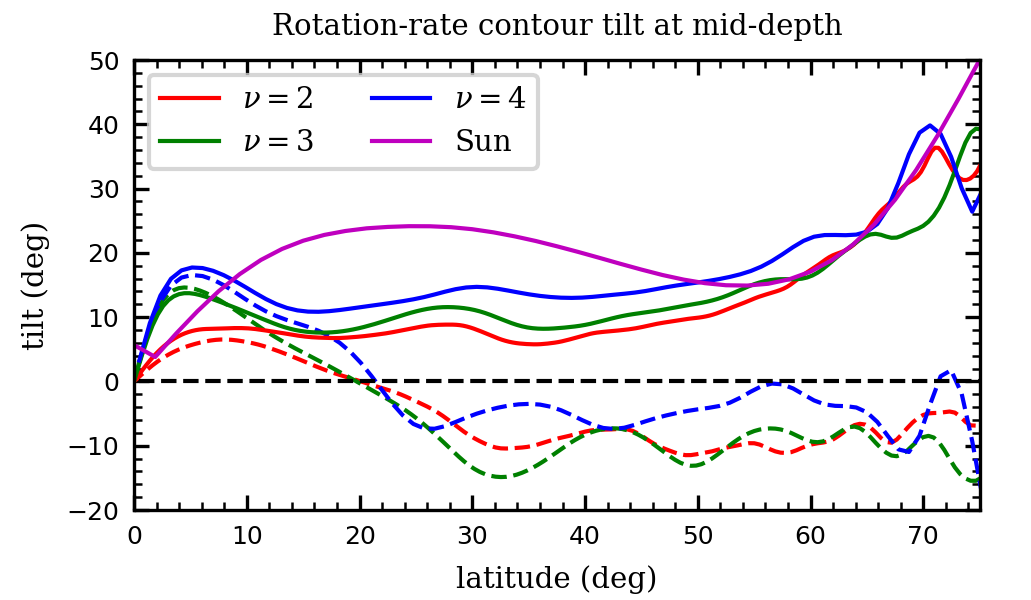}
	\caption{Tilt angle of Equation \eqref{eq:tilt} shown as a function of latitude for three of the cases rotating at $3\Omega_\odot$ and for the Sun. The profiles are taken at the middle of the shell for our models and the middle of the CZ for the Sun. The northern and southern hemispheres have been averaged assuming odd symmetry for tilt angle. Dashed lines correspond to the FE cases and solid lines to the FF cases. For the solar tilt angle, we use the inversion from GONG data 1995--2004 as reported in \citet{Howe05} and shown in Figure \ref{fig:gong}.}
	\label{fig:contour_tilt}
\end{figure}

In Figure \ref{fig:contour_tilt}, we show the values of the rotation-contour tilt angle at mid-depth for a subset of our models and for the Sun. The positive tilt for the FF cases is obvious, with the maximum tilt angle being about $+15^\circ$ for the highest value of the diffusion ($\nu=4\times10^{12}\ \rm{cm}^2\ \rm{s}^{-1}$). This is still substantially lower than the solar value for contour tilt, which attains a maximum value of $\sim$$25^\circ$ in the middle of the solar CZ. The contours in the FE cases all have positive tilt at low latitudes. At high latitudes, however, they have negative tilt, and are tilted the most (with a tilt angle of about $-10^\circ$) for the \textit{lowest} value of the diffusion ($\nu=2\times10^{12}\ \rm{cm}^2\ \rm{s}^{-1}$). 


 \section{Thermal wind balance }\label{sec:tw}
 We find that to leading order, the longitudinally and temporally averaged force balance in the meridional directions $r$ and $\theta$ (or $\lambda$ and $z$) is dominated by the Coriolis, pressure, and buoyancy forces for each simulation in this work:
 \begin{align}
 -\nabla\bigg{(}\frac{\av{P}}{\rhobar}\bigg{)} + \frac{\av{S}}{\cp}g(r)\hat{\bm{e}}_r + 2\av{v_\phi}\e_\phi\times\bm{\Omega}_0\approx 0.
 \label{eq:merforce}
 \end{align}
 Here, the angular brackets denote a combined temporal and longitudinal average. 
 

In the Earth's atmosphere, a ``thermal wind" describes a situation in which geostrophic balance (pressure balancing the Coriolis force) holds in the horizontal directions and hydrostatic balance (pressure balancing gravity) holds in the vertical direction. Equation \eqref{eq:merforce} thus represents the generalization of a thermal wind to the solar geometry, in which the vertical (radial) and horizontal extents of the flow structures are comparable (unlike in the Earth's atmosphere where the vertical extent is very small). Furthermore, the flows in the solar geometry generally have a vertical component, unlike in a classical thermal wind for which the flows are purely horizontal.
 
\begin{figure}
	\includegraphics{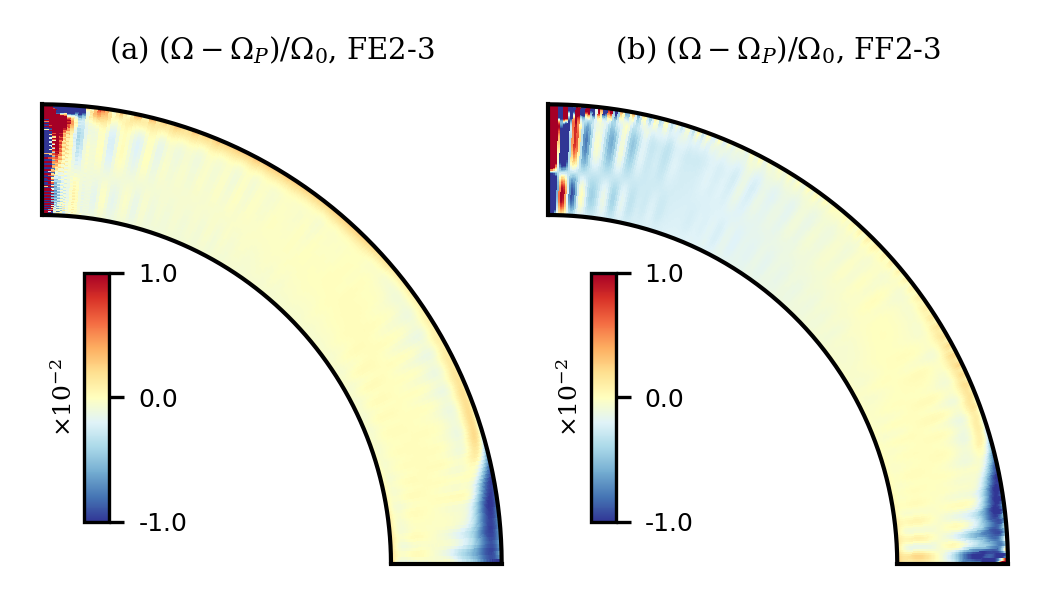}
	\caption{Point-by-point colatitudinal force balance in the meridional plane for representative pair (\textit{a}) case FE2-3 and (\textit{b}) case FF2-3. We show the difference between the temporally and longitudinally averaged rotation rate $\Omega$ and the RHS of Equation \eqref{eq:theta_force}, which we denote by $\Omega_P$ (rotation rate from the pressure), normalized by the frame rotation rate $\Omega_0$. The two hemispheres have been averaged assuming even symmetry about the equator.}
	\label{fig:theta_force}
\end{figure}
 
The colatitudinal component of Equation \eqref{eq:merforce} may be rearranged to yield
 \begin{align}
 \Omega(r,\theta) \approx \Omega_0 +  \frac{1}{\Omega_0\rhobar r^2\sin{2\theta}}\bigg{\langle}\pderiv{P}{\theta}\bigg{\rangle},
 \label{eq:theta_force}
 \end{align}
 which is a purely geostrophic equation, since the buoyancy force is radial. Figure \ref{fig:theta_force} shows a representative example of geostrophic balance for the FE2-3/FF2-3 pair. Clearly Equation \eqref{eq:theta_force} is very well satisfied for both cases, with deviations from geostrophy being no more than 1 part in $10^3$ in the bulk of the meridional plane and 1 part in $10^2$ at isolated regions by the equator and pole. The same is true for all the cases considered in this work, indicating that the differential rotation profile in our simulations is almost completely determined by the pressure profile, and vice versa. The fact that the differential rotation magnitudes are $\sim$$40\%$ greater in the FF cases compared to the FE cases is thus a consequence of greater latitudinal pressure gradients. \newtext{Figure \ref{fig:theta_force} also indicates that  viscosity plays a relatively insignificant role in the force balance at large scales.} 
 
To assess why there are opposite signs of tilt for the rotation contours in the FF and FE simulations, we differentiate Equation \eqref{eq:theta_force} with respect to the axial coordinate $z$ and use the radial component of Equation \eqref{eq:merforce} to eliminate terms (or equivalently, take the $\phi$-component of the curl of Equation \eqref{eq:merforce}), yielding
 \begin{align}
\pderiv{\Omega}{z} \approx \frac{g}{2\Omega_0r^2\sin{\theta}\cp}\bigg{\langle} \pderiv{S}{\theta}\bigg{\rangle}. 
\label{eq:domdz}
 \end{align}
 The tilt of the rotation contours is thus determined by the entropy distribution in the final thermodynamic state.
 
\begin{figure}
 	\includegraphics{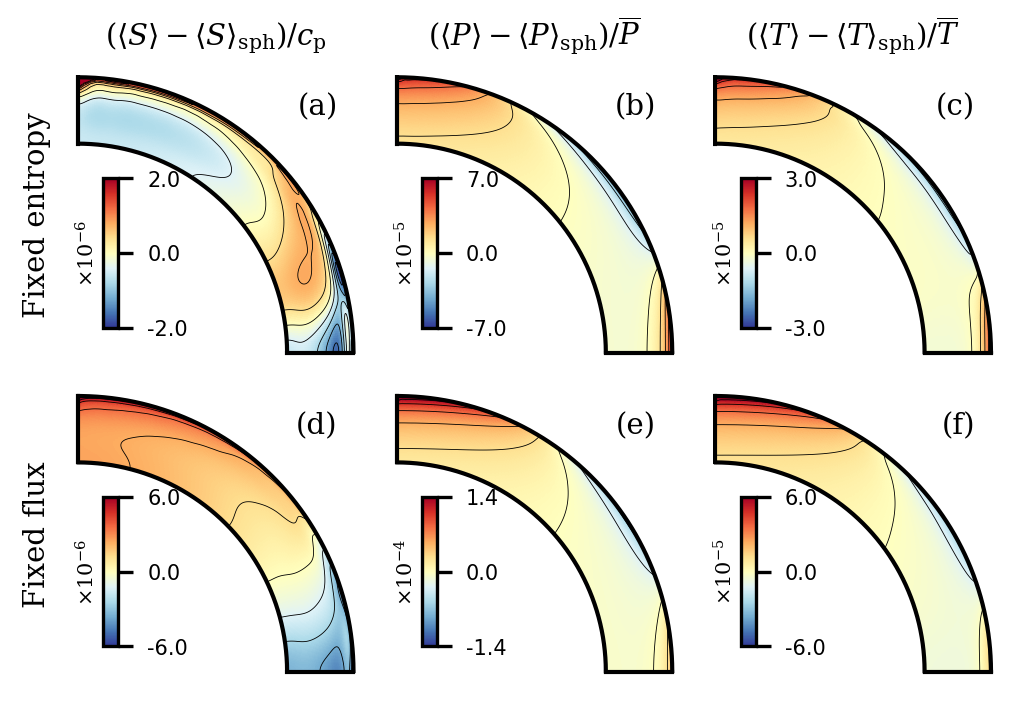}
 	\caption{Temporally and longitudinally averaged entropy, pressure, and temperature deviations from the spherically symmetric mean in the meridional half-plane (averaged assuming even symmetry about the equator) for cases FE2-3 and FF2-3, normalized by the reference state profiles. The spherical mean $\av{\cdots}_{\rm{sph}}$ has been removed to show the variation from equator to pole.}
 	\label{fig:entropy_profiles}
 \end{figure}

  In Figure \ref{fig:entropy_profiles}, we show the average profiles for entropy, pressure, and temperature in the meridional half-plane for the FE2-3/FF2-3 pair. Case FF2-3 (which is representative of all the FF cases in the simulation suite) displays a monotonically increasing entropy from equator to pole. Case FE2-3, on the other hand, has a non-monotonic entropy profile: except on the outer boundary, the entropy from equator to pole increases up to $\sim$$20^\circ$ latitude, then decreases. In both cases, the pressure and temperature deviations (normalized by the background reference state) are substantially greater (by a factor of $\sim$30 in the case of the pressure) than the entropy deviation. The profiles of pressure and temperature in the meridional plane thus tend to mirror one another, with high temperature regions corresponding to high pressure regions and vice versa (compare the last two columns of Figure \ref{fig:entropy_profiles}).
    
\begin{figure}
	\includegraphics{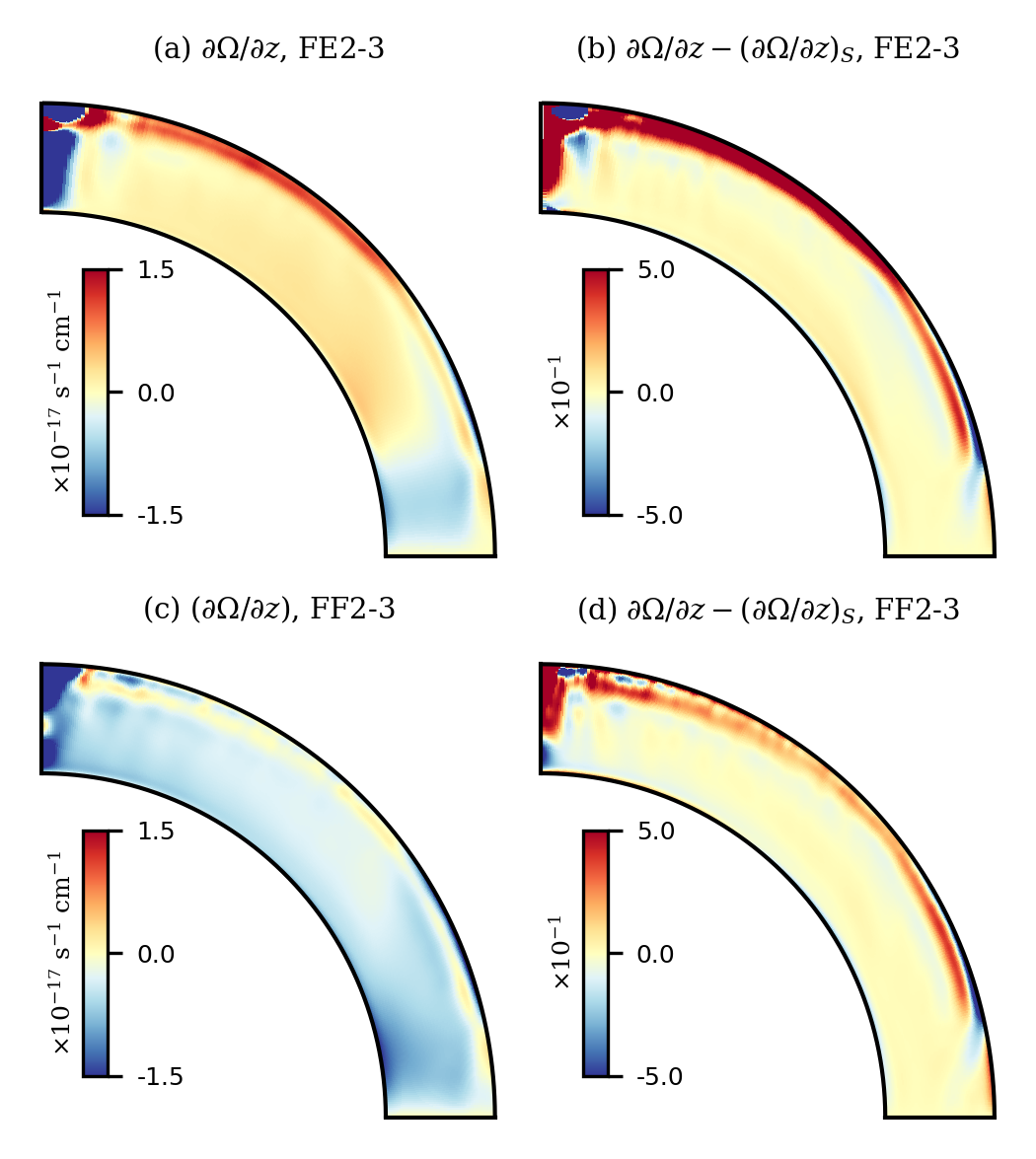}
	\caption{Temporally and longitudinally averaged azimuthal vorticity balance in the meridional plane for representative pair FE2-3 and FF2-3. The two hemispheres have been averaged assuming odd symmetry about the equator. The left-hand column (\textit{a}, \textit{c}) shows the axial derivative of the rotation rate, $\partial\Omega/\partial z$. The right-hand column (\textit{b}, \textit{d}) shows the difference between $\partial \Omega/\partial z$ and the RHS of Equation \eqref{eq:domdz} ($(\partial\Omega/\partial z)_S$, or the axial derivative of rotation rate from the entropy), normalized by $\sn{1.5}{-17}\ \rm{s^{-1}\ cm^{-1}}$.}
	\label{fig:phi_vorticity}
\end{figure}

The balance described by Equation \eqref{eq:domdz} is shown for the representative simulation pair FE2-3/FF2-3 in Figure \ref{fig:phi_vorticity}. There is good balance in the deep layers, although significant departures near the outer surface, which has been noted frequently in past work (e.g.,  \citealt{Brun11}; \citealt{Augustson12}; \citealt{Hotta15}). Quantitatively, the error in Equation \eqref{eq:domdz} (shown in the right-hand column of Figure \ref{fig:phi_vorticity}) is $\sim$10\% in the lower 80\% of the layer and $\sim$50\% in the upper 20\% of the layer. For solar-like differential rotation (fast equator and slow poles), positively-tilted rotation contours (the FF cases) correspond to $\partial\Omega/\partial z  < 0$, which arises from $\av{\partial S/\partial\theta} < 0$ at all latitudes, as in Figure \ref{fig:entropy_profiles}(\textit{d}). Similarly, the FE cases (which have contours tilted negatively at high-latitudes and positively at low latitudes) all have $\av{\partial S/\partial\theta} > 0$ at high latitudes and $\av{\partial S/\partial\theta} < 0$ at low latitudes, as in Figure \ref{fig:entropy_profiles}(\textit{a}). 

 
 \section{Poleward energy transport from Busse columns}\label{sec:en}
In light of Equations \eqref{eq:theta_force} and \eqref{eq:domdz}, a thermal wind in spherical geometry fundamentally consists of pressure and entropy differences in latitude. Poles that are high-pressure and high-entropy relative to lower latitudes (which we have shown lead to strong differential rotation with positively-tilted contours) are expected to be established by the preferentially poleward transport of energy. In our simulations, this transport arises from the action of the convective Busse-column rolls. These rolls manifest at convective onset as an overstable, low-frequency prograde wave (e.g., \citealt{Unno89}) or, as it is called in the geophysics literature, a thermal Rossby wave. This wave consists of a series of convective rolls, or Busse columns, that gird the equator. Each roll is rotationally-aligned and the sign of the vorticity alternates from roll to roll. Furthermore, each roll is in geostrophic balance; hence, the alternating sign of the vorticity corresponds to every other roll being a zone of high pressure, with low-pressure rolls in between. Since the ends of the columns (at mid-latitudes) have neutral pressure, the pressure anomalies at the equator cause poleward axial flow in the high-pressure rolls and equatorward flow in the low-pressure rolls (e.g., Figure 1 in \citealt{Gilman83}). The resulting strong correlation between pressure and the direction of axial flow leads to a net poleward enthalpy flux through pressure work. 

\begin{figure*}
	\includegraphics{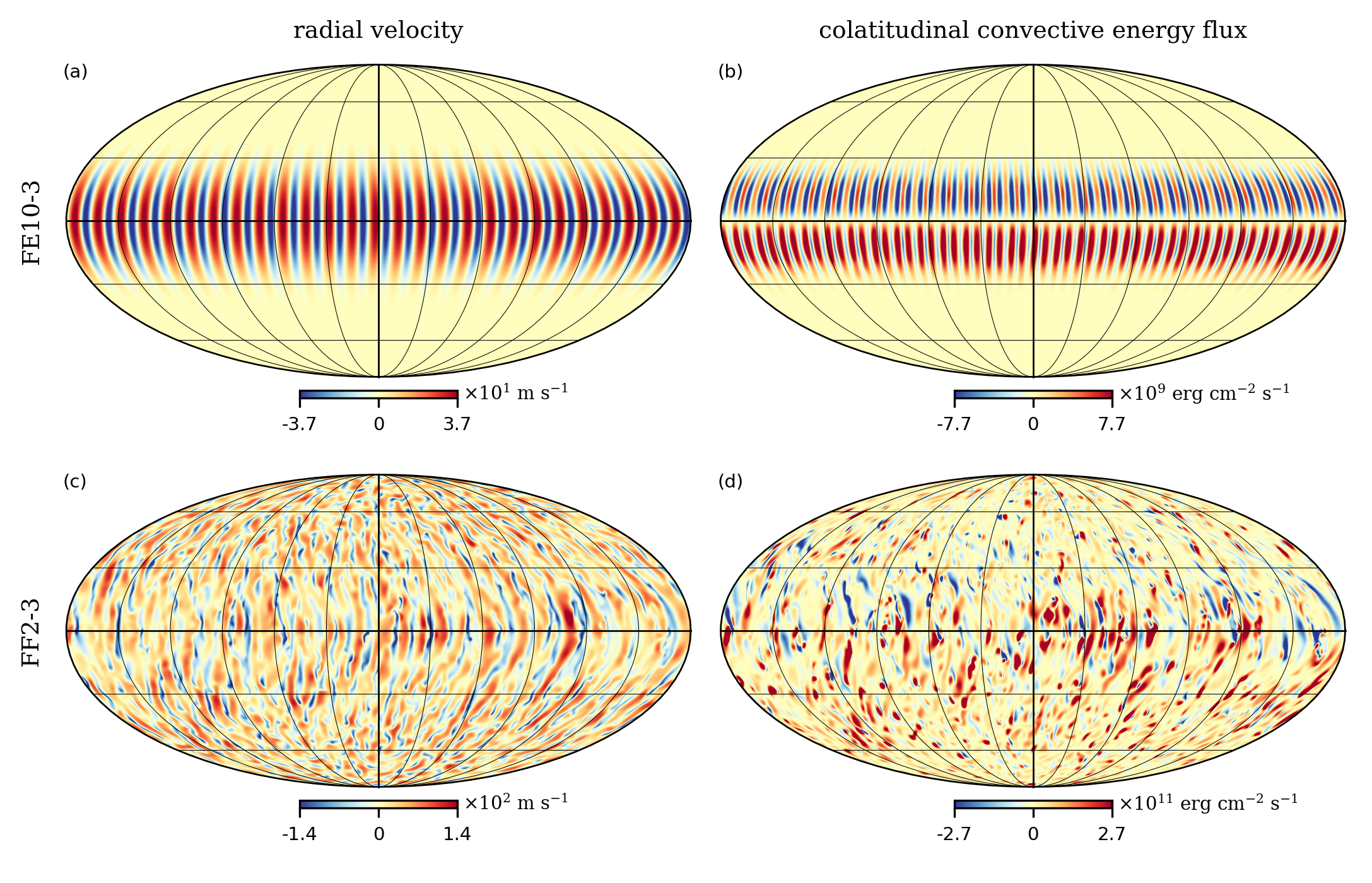}
	\caption{Spherical snapshots (at $r/\ro=0.910$) of the radial velocity $v_r^\prime$ (lefthand panels) and colatitudinal convective energy flux $v_\theta^\prime [\overline{T}S^\prime + P^\prime + (1/2)\rhobar (v^\prime)^2]$ (righthand panels), shown in global Mollweide projections. The three terms in the convective energy flux correspond to advection of heat, pressure work, and advection of kinetic energy, respectively. The top row of panels is taken from the highly diffusive, barely supercritical case FE10-3 and the bottom row of panels is taken from case FF2-3. \newtext{In both cases, the Busse-column sites in the lefthand panels are associated with sites of poleward convective energy flux in the righthand panels---i.e., panels (\textit{b}) and (\textit{d}) are on average blue in the north and red in  the south.}}
	\label{fig:sslice}
\end{figure*}

The effect just described is easiest to illustrate for models that are barely supercritical. Here, the profiles for the velocity and thermodynamic variables are dominated by the wavenumber associated with the most unstable mode. For the range of Ekman numbers spanned by our simulation suite, the resulting Busse columns are mostly localized in the outer half of the shell by radius and at low latitudes (see \citealt{Jones09} for a linear stability analysis of the problem). Figures \ref{fig:sslice}(\textit{a}, \textit{b}) show the instantaneous convective radial velocity and convective colatitudinal energy transport in the highly diffusive case FE10-3, which lies in the barely supercritical regime. Each upflow and downflow (pairs of which trace one Busse column roll) has an associated colatitudinal energy transport that is, on average, negative in the northern hemisphere and positive in the southern hemisphere, implying preferentially poleward energy transport.   \newtext{Note that under the spherical-coordinate convention, with $\theta$ as the colatitude, the positive-$\theta$ direction is always oriented north-to-south. Thus equatorward (poleward) transport of energy corresponds to positive (negative) colatitudinal transport in the northern hemisphere and negative (positive) transport in the southern hemisphere.}

Figures \ref{fig:sslice}(\textit{c}, \textit{d}) show the radial velocity and convective energy transport in the comparatively more turbulent case FF2-3. The flow structures are more intricate and fine-scale than in the barely supercritical regime, but the imprint of the most unstable mode remains. \newtext{Many Busse column rolls---which can be seen at low latitudes as columnar red and blue features in Figure \ref{fig:sslice}(\textit{c})---correspond to sites of negative colatitudinal energy transport (blue in Figure \ref{fig:sslice}(\textit{d})) in the northern hemisphere and positive transport  (red in Figure \ref{fig:sslice}(\textit{d})) in the southern hemisphere. Overall, there are more sites of poleward energy transport (from the Busse columns) than sites of equatorward transport in each hemisphere. Under an azimuthal average, the Busse columns in the more turbulent case FF2-3 thus yield preferentially poleward energy transport, just as in the barely supercritical case FE10-3.}

\begin{figure*}
	\includegraphics{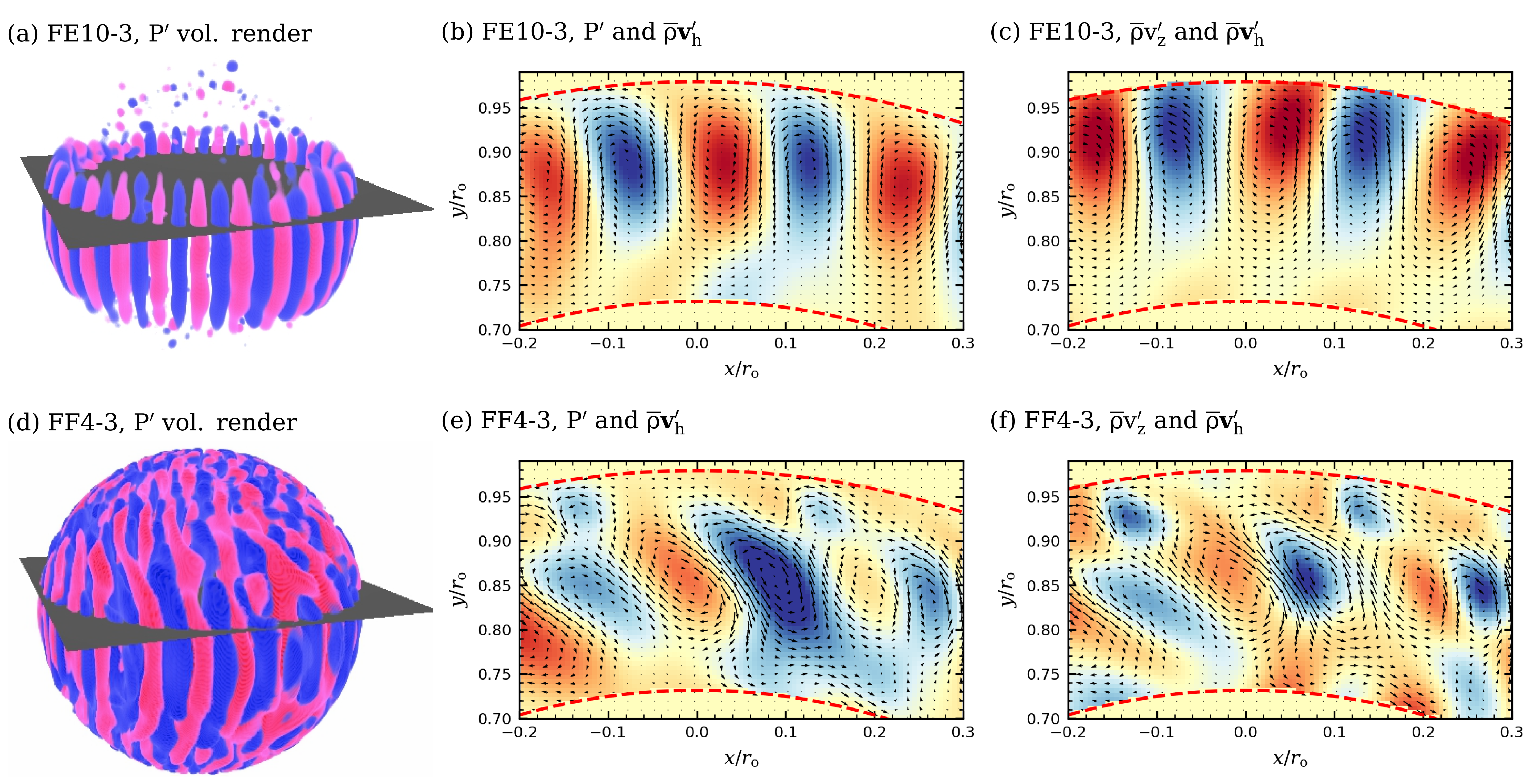}
	\caption{Geostrophic balance in Busse columns for cases FE10-3 (top row) and FF4-3 (bottom row). Here, we use the standard Cartesian coordinates $x$, $y$, and $z$.  (\textit{a} and \textit{d}) Snapshots of the non-axisymmetric pressure ($P^\prime = P - \av{P}$) shown as 3D volume renderings, with the view from slightly north of the equator and the scene cut by the plane $z/\ro=0.2$. (\textit{b} and \textit{e}) Closeup views of $P^\prime$ (shown in color) and $\rhobar\bm{v}_{\rm{h}}^\prime \equiv \rhobar(v_x^\prime\e_x + v_y^\prime\e_y)$ (shown as a vector field) in a portion of the plane $z/\ro=0.2$. (\textit{c} and \textit{f}) Closeup views of $\rhobar v_z^\prime$ and $\rhobar\bm{v}_{\rm{h}}^\prime$ in the same portion of the plane as in (\textit{b} and \textit{e}). In all panels, red tones indicate positive values and blue tones indicate negative values. }
	\label{fig:column_cuts}
\end{figure*}

The geostrophic nature of the Busse columns is illustrated in Figure \ref{fig:column_cuts}, as is the resulting axial component of the flow. In the top row (case FE10-3), panel \textit{a} shows that the Busse-column rolls alternate between high and low pressure. Panel \textit{b} shows that the high-pressure rolls are each anticyclonic (have negative vorticity), while the low-pressure rolls are cyclonic. Finally, panel \textit{c} shows that each high-pressure anomaly corresponds to poleward flow ($v_z > 0$ in the northern hemisphere), while each low-pressure anomaly corresponds to equatorward flow ($v_z<0$). In the bottom row (the more supercritical case FF4-3), the Busse columns are less regularly spaced, but still largely alternate between anticyclonic regions of high pressure  and cyclonic regions of low pressure (panels \textit{d}, \textit{e}). The axial flow associated with the Busse columns in case FF4-3 (panel \textit{f}) then leads to poleward energy transport through pressure work, just as in case FE10-3.

\newtext{It has long been known that Busse columns transport angular momentum outward. We have just shown that Busse columns also transport heat poleward. The Busse columns thus define a purely hydrodynamic mechanism coupling entropy and angular velocity. \citet{Balbus09b} posited the presence of such a convective mechanism in the Sun and further argued that the motions responsible should fall along surfaces of both constant entropy and constant angular velocity. In that picture, the isorotational and isentropic contours should thus coincide. The Busse columns in our simulation suite do not completely behave in this way, as evidenced by none of our simulations having good alignment of the isorotational  and isentropic contours. Independent of whether the constant-entropy and constant-angular-velocity surfaces coincide in the Sun, a key point from our work is that the Busse columns provide an explicit convective mechanism to couple entropy and angular velocity.}

\section{Effect of outer thermal boundary condition}\label{sec:bc}
Given that Busse columns direct energy poleward, equilibrium can be achieved by forming conductive gradients that balance the poleward convective enthalpy flux. In general, such conductive transport can be achieved in two distinct ways. As the pole heats up and the equator-to-pole contrast increases, a latitudinal gradient will form that transports heat equatorward. Additionally, the increased temperature of the pole can lead to enhancement of the radial gradients in the outer thermal boundary layer, thus causing the poles to lose heat more efficiently (i.e., become superluminous).  In the FF cases, the outer thermal boundary condition precludes the second of these options because the radial gradients are fixed. Hence, the FF models must rely solely on the development of a pole-to-equator conductive gradient. In the FE models, both types of gradients are possible. Therefore, the amount that the pole must be heated before equilibrium can be achieved is less for the FE models than it is for the FF models. The outer thermal boundary condition thus has a direct influence on the latitudinal  contrast in the temperature, entropy, and pressure, with the FF boundary condition being conducive to strong contrast in all the thermodynamic variables. In the presence of thermal-wind balance, the FF boundary condition thus leads to enhanced contrast in the differential rotation and positively-tilted isocontours in the rotation rate.

Mathematically, we illustrate the combined effects of the outer thermal boundary condition and latitudinal energy transport using the steady-state total energy equation for the fluid. Using Equations \eqref{eq:cont}--\eqref{eq:en}, this equation takes the form of a balance of fluxes:
\begin{align}
\nabla\cdot\bm{\mathcal{F}} = 0, 
\label{eq:toten}
\end{align}
where 
\begin{align}
\f\equiv \f_{\rm{conv}} +\f_{\rm{cond}} + \f_{\rm{rad}} + \f_{\rm{visc}} + \f_{\rm{circ}}
\label{eq:totflux}
\end{align}
is the temporally and longitudinally averaged total energy flux in the meridional plane and we have defined the averaged convective, conductive, radiative, viscous, and meridional-circulation fluxes through
\begin{subequations}\label{def:fluxes}
	\begin{align}
	\f_{\rm{conv}} &\equiv \rhobar\bigg{(}\cp \av{T^\prime\bm{v}^\prime} + \frac{1}{2}\av{v^2\bm{v}}\bigg{)},\\
	\f_{\rm{cond}} &\equiv -\kappa\rhobar\overline{T}\av{\nabla S},\\
	\f_{\rm{rad}} &\equiv \Bigg{(}\frac{1}{r^2}\int_r^{r_o}Q(x)x^2dx\Bigg{)}\hat{\bm{e}}_r,\\
    \f_{\rm{visc}} &\equiv  -\av{\bm{D}\cdot\bm{v}}, \\
	\text{and}\ \ \ \ \ \f_{\rm{circ}} &\equiv  \rhobar\cp \av{T}\av{\bm{v}}
	\end{align}
\end{subequations}
respectively. Note that $\rhobar\overline{T} S + P =\cp T$, so the terms with $\av{T^\prime\bm{v}^\prime}$ and $\av{T}\av{\bm{v}}$ in the convective and meridional-circulation fluxes represent the combined effects of heat advection and pressure work. Technically, the flux due to transport of kinetic energy (proportional to $\av{v^2\bm{v}}$) has convective terms (e.g., $\av{(v^\prime)^2 \bm{v}^\prime}$) \textit{and} meridional-circulation terms (e.g., $\av{\bm{v}}^2\av{\bm{v}}$). For simplicity, we include all the kinetic-energy terms in the convective flux since they are in general small. 
 
We are interested in the total latitudinal transport of energy, and so we integrate the total flux in Equation \eqref{eq:totflux} over conical surfaces at constant latitude:
 \begin{align}
 \I_\theta(\theta) \equiv 2\pi\sin{\theta} \int_{\ri}^{\ro}\mathcal{F}_\theta(r,\theta)rdr.
 \end{align}

For the FF cases, there can be no net transport of energy in latitude due to the absence of conductive losses in the polar regions through the outer boundary. In other words, $\I_\theta(\theta) \equiv 0$. For the FE cases, by contrast, there is a net poleward energy transport because the poles are allowed to be superluminous. Thus, $\I_\theta(\theta)$ will in general be negative in the northern hemisphere and positive in the southern hemisphere.  \newtext{Recall that equatorward (poleward) transport of energy corresponds to positive (negative) $\I_\theta(\theta)$ in the northern hemisphere and negative (positive) $\I_\theta(\theta)$ in the southern hemisphere.}

\begin{figure}
  	\includegraphics{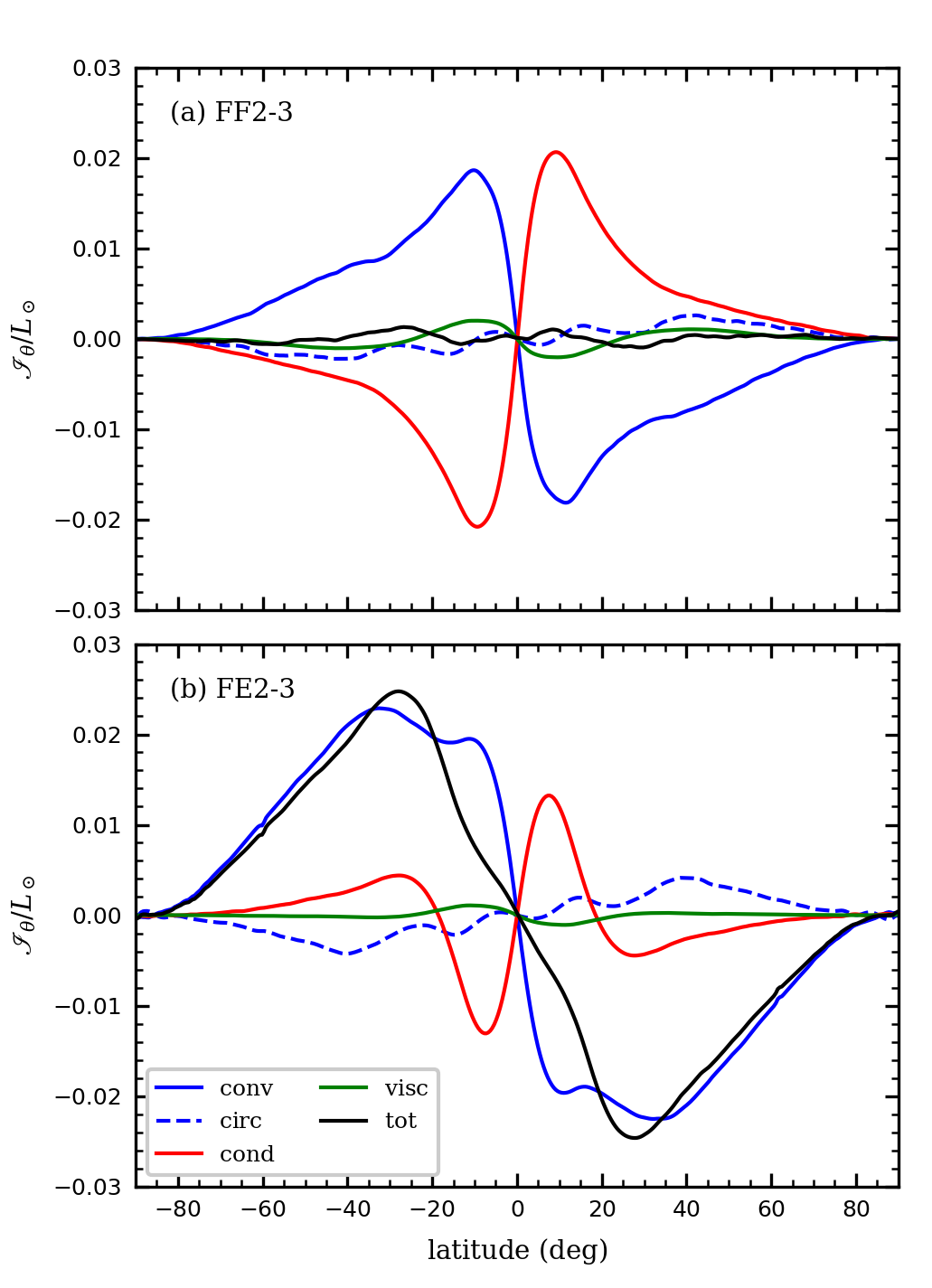}
  	\caption{Total colatitudinal energy transport for (\textit{a}) case FF2-3 and (\textit{b}) case FE2-3, or integrals of the fluxes in Equation \eqref{def:fluxes} over conical surfaces at constant latitude. \newtext{The integrated fluxes are plotted as functions of latitude ($=\pi/2 - \theta$).} Contributions from the various fluxes are indicated in the legend, ``tot" denoting the sum of all the other fluxes. \newtext{Positive (negative) $\I_\theta$ indicates equatorward (poleward) transport in the northern hemisphere and poleward (equatorward) transport in the southern hemisphere.}}
  	\label{fig:eft}
  \end{figure}

 Figure \ref{fig:eft}(\textit{a}) shows the integrated colatitudinal energy flux in case FF2-3 after the system has achieved statistical equilibrium. The total flux is very close to zero at all latitudes, indicating a well-equilibrated state. The dominant transport components are the convective flux, which transports energy preferentially poleward due to the Busse columns, and the conductive flux, which transports energy equatorward. The monotonic entropy gradient of Figure \ref{fig:entropy_profiles}(\textit{d}), and by extension radially tilted contours in the FF cases, is thus seen to be a result of the response by conduction to the convective transport of energy to the poles. 
  
Figure \ref{fig:eft}(\textit{b}) shows the integrated colatitudinal energy transport in case FE2-3. The poles are clearly superluminous---i.e., there is a net poleward energy transport due to the convection. For all the FE cases explored here, the energy loss at the poles is even greater than the heating by the convection; the conductive flux is thus forced to change sign at mid-latitudes (around $\pm20^\circ$), transporting energy poleward in concert with the Busse columns. This results in the non-monotonic entropy profile of Figure \ref{fig:entropy_profiles}(\textit{a}), leading to the tilts of the rotation contours being negative at high latitudes and positive at low latitudes, as in the top row of Figure \ref{fig:diffrot_halfplane}.
   
 \section{Discussion and conclusions}\label{sec:disc}
  We have shown that the differential rotation achieved in global, 3D convection simulations is well-described by a thermal wind and highly sensitive to the outer thermal boundary condition. The FF boundary tends to yield more solar-like rotation profiles (strong contrast with positively-tilted contours), while the FE boundary yields weaker contrast and negatively-tilted contours. In light of these results, we now discuss the likelihood that the Sun's strong rotation contrast and positively-tilted contours arise from thermal-wind balance in the deep interior coupled with the observation that the radiative flux from the solar photosphere does not vary appreciably with latitude. 
  
\begin{figure}
 	\makebox[\linewidth][c]{\includegraphics[width=2in]{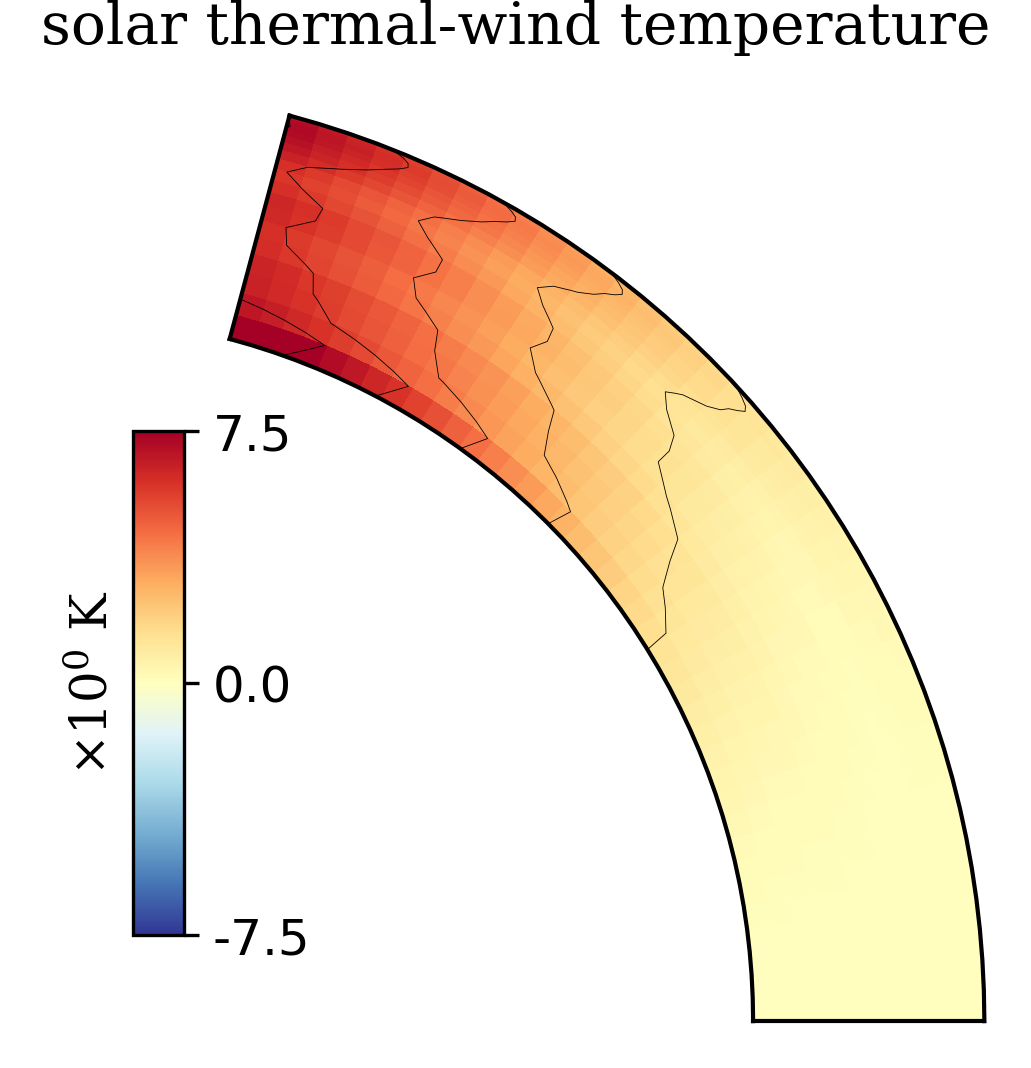}}
 	\caption{Temperature deviation from the spherical mean in the solar CZ, assuming thermal-wind balance holds. The temperature has been calculated from the rotation rate inferred from helioseismology (Figure \ref{fig:gong}), using  Equations \eqref{eq:theta_force}, \eqref{eq:domdz}  and\eqref{eq:eos}.}
 	\label{fig:solarT}
 \end{figure}

 The first question is whether the force balance Equation \eqref{eq:merforce}, which should in general be true for low Rossby numbers, holds in the Sun. The interior solar Rossby number is currently unknown, but recent helioseismic estimates \citep{Hanasoge12, Greer15} give $\rm{Ro}\lesssim0.1$ in the deep interior. Thus, it is likely that Equation \eqref{eq:merforce} (and the derivative thermal wind Equations \eqref{eq:theta_force} and \eqref{eq:domdz}) apply in the solar CZ, except perhaps in the layers just below the photosphere. Thermal-wind balance in the Sun is also not prohibited by observational results. \newtext{We can derive the temperature in the solar CZ from the rotation rate of Figure \ref{fig:gong} assuming thermal-wind balance holds, integrating Equations \eqref{eq:theta_force} and \eqref{eq:domdz}  to get $P$ and $S$ in the meridional plane and using Equation \eqref{eq:eos} to get $T$. In this calculation, we set $\Omega_0$ to the sidereal Carrington rate and use the same polytropic reference state employed in our models for the solar profiles.  The result is shown in Figure \ref{fig:solarT}. The equator-to-pole temperature contrast required to drive the solar-like differential rotation is $\sim$10 K (approximately uniform with radius), which is well below the detection limit of helioseismology (e.g., \citealt{Brun10}).}
 
  \begin{figure}
 	\includegraphics{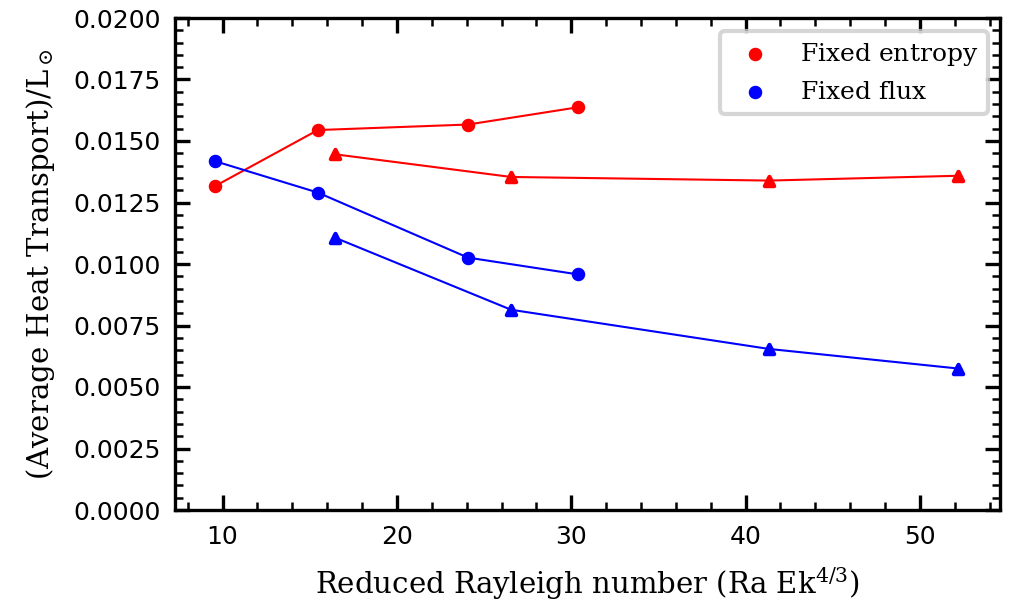}
 	\caption{Net poleward transport of heat from Busse columns for all simulations in the suite (except the barely supercritical cases FE10-3 and FF10-3) plotted versus the reduced Rayleigh number $\mathcal{R} \equiv\rm{Ra\ Ek^{4/3}}$. The poleward heat flux $|\mathscr{I}_{{\theta,\ \rm{conv}}}|$ has been averaged over latitude. Circles mark the $\Omega_0=3\Omega_\odot$ cases and triangles the $\Omega=2\Omega_\odot$ cases.}
 	\label{fig:av_heatflux}
 \end{figure}
 
 The second question is whether the Sun's Busse columns send energy preferentially poleward. In general, stellar convection transitions through a series of convective regimes as the supercriticality (measured by the reduced Rayleigh number $\mathcal{R}$) increases \citep{Hindman20}. Both the least supercritical case in our work (FE10-3, for which $\mathcal{R} \sim 2$) and the most supercritical cases (the pair FF2-2 and FE2-2, for which $\mathcal{R} \sim50$) have a strong preference for poleward transport by Busse columns, suggesting that the poleward transport is a feature of the most unstable mode of convection that stays present as the flows get ever more complex. 
 
 Finally, it is an open question how the Sun might transport energy equatorward to maintain equilibrium. In our simulations, the net poleward transport of energy by Busse columns is at its maximum a few per cent of the solar luminosity, which is counteracted almost entirely by conduction in the FF cases (Figure \ref{fig:eft}). In the Sun, the thermal diffusivity associated with radiative heating is $\sim$$10^7\ \rm{cm^2\ s^{-1}}$ at mid-CZ (e.g., \citealt{Hindman20}), which (if a thermal wind were operating with a temperature contrast of $\sim$10 K) would correspond to a latitudinal energy flux of $\sim$$10^{-7} L_\odot$.  \newtext{Figure \ref{fig:av_heatflux} shows how the Busse-column heat flux scales with supercriticality. The trends are different between the FE and FF cases (and the $\Omega_0=3\Omega_\odot$ and $\Omega_0=2\Omega_\odot$ cases), but there is clearly a tendency for the net transport to go down for our more supercritical FF simulations (blue curves). This indicates that the balance between conductive and convective heat flux could hold in the Sun, just with much smaller flux magnitudes. We admit that these results are only suggestive, since the flux in all our models is still orders of magnitude higher than the presumed solar value of $10^{-7}L_\odot$.}

 

\begin{figure}
 	\includegraphics{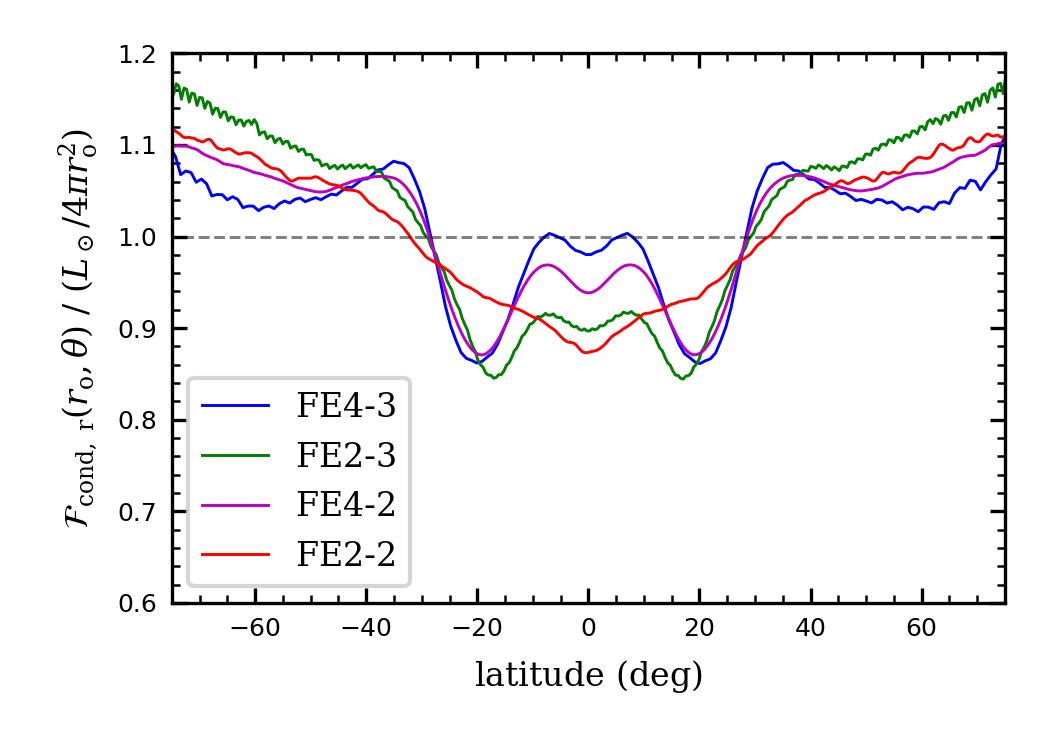}
 	\caption{Latitudinal profile of $\mathcal{F}_{\rm{cond},\ r}$ at the outer boundary for several of the FE cases, normalized by the flux needed to carry out the solar luminosity equally at all latitudes. The flux has been averaged over time and longitude.}
 	\label{fig:fcondr_vs_theta_top}
 \end{figure}

 On a more practical note, it is advantageous to use an FF outer boundary condition in solar simulations for two reasons. First, maintaining a strong differential rotation is particularly relevant for dynamo models, since the dynamo cannot be maintained without sufficient shear (e.g., \citealt{Brown10}; \citealt{Guerrero16}; \citealt{Matilsky20}; \citealt{Bice20}). Second, using an FE outer boundary condition leads to superluminous poles, which are directly at odds with solar observations. Figure \ref{fig:fcondr_vs_theta_top} shows the conductive flux as a function of latitude at the top of the domain for the FE cases. For case FE2-3, the flux in the polar regions reaches a value in excess of the solar luminosity by about 20\%. This is far greater than the observationally-constrained value of $<1\%$ for the Sun \citep{Rast08}. 

 We very much view this paper as a complement to \citet{Miesch06}. In that work, a systematic tilt of the rotation contours was achieved by imposing a small latitudinal entropy gradient at the inner boundary, \newtext{thereby ensuring that the entropy increased monotonically from equator to pole}. And indeed, for all our FF cases, there is a similar \newtext{monotonic equator-to-pole} entropy gradient at the inner boundary. \newtext{Monotonicity is \textit{not} achieved } in the FE cases. In other words, \citet{Miesch06} showed that it is possible to tilt the rotation contours by imposing a preferred geostrophic balance, and here we have shown how this preferred balance is naturally established as the result of poleward energy transport by Busse columns and the FF outer boundary condition. 
 
\newtext{ We conclude that any successful dynamical model of the Sun's convection zone must be consistent with three key observations: (1) the Sun possesses strong latitudinal differential rotation, (2) the isorotation contours are tilted positively in the sense of Figure \ref{fig:schematic_tilt}, and (3) the emergent intensity is nearly spherically symmetric. We have demonstrated in this paper that the transport of angular momentum and heat by Busse columns can yield results consistent with observations (1) and (2).  Busse columns transport angular momentum outward, away from the rotation axis, thus speeding up the equator. They simultaneously transport heat poleward via the enthalpy flux associated with pressure work. These two transports reach equilibrium once thermal-wind balance is achieved.}
 
 \newtext{While it is satisfying to address the observations (1) and (2) with a single mechanism (i.e., the action of Busse columns), observation (3)  presents a theoretical problem. The  thermal-wind balance resulting from Busse columns requires that the solar poles be at a higher temperature than the equator, with a contrast of $\sim$10 K throughout the CZ. The emergent intensity at the solar surface, however, is nearly uniform in latitude, with less than $\sim$2.5 K variation in effective temperature. This suggests that the thermal-wind balance must only be maintained in the deep CZ, and the surface layers (perhaps within the near-surface shear layer) may have different dynamics. Those dynamics must somehow screen the surface from the imposition of the latitudinal temperature gradient from below.}
 
 \newtext{In our models, the proper proxy for the emergent intensity is the radial conductive flux at the outer surface. For the FF cases, we impose spherical symmetry of this flux as a boundary condition. Further, we have shown  that this boundary condition is conducive to achieving strong differential rotation and properly tilted rotational isocontours. However, a fully self-consistent model of the solar interior would reproduce the spherical symmetry at the outer boundary as a natural consequence of the near-surface convection and not as an  enforced boundary condition.}
 
 \newtext{Since the underlying dynamics giving rise to latitudinally-independent emissivity in the Sun are not well-understood, we cannot predict a priori whether that same mechnism would be sufficient to enforce spherical symmetry for other stars. It has long been postulated that oblateness induced by rotation in high-mass stars (in which a radiation layer overlies an interior CZ) could result in superluminous poles (e.g., \citealt{vonZeipel24}; \citealt{Collins63}). Here, we have identified Busse columns as a possible mechanism for producing hot poles in low-mass stars (in which the CZ overlies the radiation layer) that does not depend on the star's distortion. Furthermore, Busse columns do not necessarily require rapid rotation---they simply require strong rotational constraint, or in other words, low Rossby number. Thus, small dim stars that are expected to have weak convection may have active Busse columns and hot poles, even with only moderate rotation. Finally, since the homogeneity (or lack thereof) of the stellar emissive flux may significantly affect the interior differential rotation, we should not assume that the positive-tilts of the rotational isocontours inferred for the Sun are a general feature of other low-mass stars.}
  \section*{Appendix: Diagnostic Parameters for the Simulation Suite}\nonumber\label{ap:tab2}
  \newtext{Table \ref{tab:output} contains various input and diagnostic parameters for the entire simulation suite as defined in Section \ref{sec:res}. Recall that for each simulation name, ``FE" or ``FF" refers to the type of outer thermal boundary condition employed, the first number gives the diffusion value $\nu=\kappa$ in units of $10^{12}\ \rm{cm^2\ s^{-1}}$, and the second number after the dash gives the frame rotation rate $\Omega_0$ in units of $\Omega_\odot\equiv \sn{2.87}{-6}\ \rm{rad\ s^{-1}}$, or $\Omega_\odot/2\pi = 456\ \rm{nHz}$. $N_r$, $N_\theta$, and $N_\phi$ refer to the number of radial, colatitudinal, and longitudinal grid points, respectively. The run time is given in units of the thermal diffusion time, defined to be $H^2/\kappa$, with $H$ the shell depth.}
 \begin{table*}
 	\centering
 	\caption{Fluid diagnostic parameters for the different simulations.}\label{tab:output}
 	\begin{tabular}{c  c  c c  c  c c c c}
 		\hline\hline
 		Name & $(N_r, N_\theta, N_\phi)$ & $\raf$ & Re = Pe & Ek & Ro &$\mathcal{R} \equiv\rm{Ra\ Ek^{4/3}}$ & $\Delta\Omega/\Omega_0$ & Run time\\
 		\hline
 		FE2-3 &  (128, 384, 768) & $\sn{8.52}{5}$ & 63.2 & $\sn{4.62}{-4}$ & $\sn{3.38}{-2}$ & 30.4 & 0.097 & 23.6 \\ 
 		FE2.3-3 &  (128, 384, 768) & $\sn{5.60}{5}$ & 51.8 & $\sn{5.31}{-4}$ & $\sn{3.21}{-2}$ & 24.1 & 0.092 & 7.77 \\ 
 		FE3-3 & (96, 384, 768) & $\sn{2.52}{5}$ & 35.2 & $\sn{6.93}{-4}$ & $\sn{2.90}{-2}$ & 15.5 & 0.078 & 59.6 \\	
 		FE4-3 & (128, 192, 384) & $\sn{1.07}{5}$ & 21.9 & $\sn{9.23}{-4}$ & $\sn{2.55}{-2}$ & 9.62 & 0.062 & 33.4 \\
 		FE10-3 & (64, 96, 192) & $\sn{6.82}{3}$ & 2.54 & $\sn{2.31}{-3}$ & $\sn{7.45}{-3}$ & 2.08 & 0.0048 & 238.5\\
 		\hline
 		FE2-2 & (128, 576, 1152) &  $\sn{8.52}{5}$ & 75.6 & $\sn{6.93}{-4}$ & $\sn{5.91}{-2}$ & 52.2 & 0.082 & 7.78\\ 		
 		FE2.3-2 & (128, 576, 1152) &  $\sn{5.60}{5}$ & 62.2 & $\sn{7.96}{-4}$ & $\sn{5.61}{-2}$ & 41.3 & 0.088 & 9.14\\
 		FE3-2 & (96, 384, 768) &  $\sn{2.52}{5}$ & 42.8 & $\sn{1.04}{-3}$ & $\sn{5.09}{-2}$ & 26.6 & 0.092 & 11.9\\
 		FE4-2 & (64, 288, 576) &  $\sn{1.07}{5}$ & 28.0 & $\sn{1.39}{-3}$ & $\sn{4.51}{-2}$ & 16.6 & 0.086 & 47.7\\
 		\hline\hline
 		FF2-3 & (128, 576, 1152) & $\sn{8.52}{5}$ & 64.5 & $\sn{4.62}{-4}$ & $\sn{3.45}{-2}$ & 30.4 & 0.140 & 14.7 \\
 		FF2.3-3 &  (128, 384, 768) & $\sn{5.60}{5}$ & 52.9 & $\sn{5.31}{-4}$ & $\sn{3.28}{-2}$ & 24.1 & 0.131 & 7.90 \\ 
 		FF3-3 & (96, 384, 768) & $\sn{2.52}{5}$ & 35.8 & $\sn{6.93}{-4}$  & $\sn{2.95}{-2}$ & 15.5 & 0.108 & 14.4 \\		 			 		
 		FF4-3 & (128, 192, 384) & $\sn{1.07}{5}$ & 22.4 & $\sn{9.23}{-4}$ & $\sn{2.46}{-2}$ & 9.62 &0.080 & 25.5\\ 
 		FF10-3 & (64, 96, 192) & $\sn{6.82}{3}$ & 2.89 & $\sn{2.31}{-3}$ & $\sn{8.58}{-3}$ & 2.08 & 0.0024 & 15.9\\
 		\hline
 		FF2-2 & (128, 576, 1152) & $\sn{8.52}{5}$ & 81.2 & $\sn{6.93}{-4}$ & $\sn{6.33}{-2}$ & 52.2 & 0.110 & 6.68 \\ 		
 		FF2.3-2 & (128, 576, 1152) & $\sn{5.60}{5}$ & 65.3 & $\sn{7.96}{-4}$ & $\sn{5.89}{-2}$ & 41.3 & 0.121 & 8.27 \\ 		
 		FF3-2 & (96, 384, 768) & $\sn{2.52}{5}$ & 44.4 & $\sn{1.04}{-3}$ & $\sn{5.30}{-2}$ & 26.6 & 0.128 & 10.3 \\
 		FF4-2 & (128, 384, 768) & $\sn{1.07}{5}$ & 28.9 & $\sn{1.39}{-3}$ & $\sn{4.66}{-2}$ & 16.6 & 0.117 & 70.4 \\
 		\hline
 	\end{tabular}
 	\tablecomments{The 18 simulations here are in four groupings according to FE or FF at two different rotation rates.}
 \end{table*}

\acknowledgments
We thank Mark Miesch for elucidating how poleward energy transport might arise from Busse columns and Rachel Howe for providing invaluable helioseismic data. We thank the anonymous reviewer for providing detailed feedback that resulted in a significantly improved manuscript. Loren Matilsky was partly supported during this work by the Future Investigators in NASA Earth and Space Sciences Technology (FINESST) award 80NSSC19K1428 and by a George Ellery Hale Graduate Fellowship. This research was primarily supported by NASA Heliophysics through grants NNX13AG18G, NNX17AG22G, 80NSSC18K1127,  80NSSC17K0008, 80NSSC18K1125, and 80NSSC19K0267. Computational resources were provided by the NASA High-End Computing (HEC) Program through the NASA Advanced Supercomputing (NAS) Division at Ames Research Center. Rayleigh has been developed by Nicholas Featherstone with support by the National Science Foundation through the Computational Infrastructure for Geodynamics (CIG) under NSF grants NSF-0949446 and NSF-1550901.


\clearpage

\begin{thebibliography}{}
\bibitem[Anders et al. (2020)]{Anders20} Anders, E.H., Vasil, G.M., Brown, B.P., \& Korre, L. 2020, PhRvF, submitted, \MYhref{https://arxiv.org/abs/2003.00026}{arXiv:2003.00026}

\bibitem[Augustson et al. (2012)]{Augustson12} Augustson, K.C., Brown, B.P., Brun, A.S.,  Miesch, M.S., \& Toomre, J. 2012,
\href{https://doi.org/10.1088/0004-637X/756/2/169}{\apj}, \MYhref{https://ui.adsabs.harvard.edu/abs/2012ApJ...756..169A/abstract}{756, 169}


\bibitem[Balbus (2009a)]{Balbus09a} Balbus, S.A. 2009, 
\href{https://doi.org/10.1111/j.1365-2966.2009.14469.x}{\mnras}, 
\MYhref{https://ui.adsabs.harvard.edu/abs/2009MNRAS.395.2056B/abstract}{395, 2056}

\bibitem[Balbus et al. (2009b)]{Balbus09b} Balbus, S.A. Bonart, J., Latter, H.N., \& Weiss, N.O. 2009, 
\href{https://doi.org/10.1111/j.1365-2966.2009.15464.x}{\mnras}, 
\MYhref{https://ui.adsabs.harvard.edu/abs/2009MNRAS.400..176B/abstract}{400, 176}

\bibitem[Bice \& Toomre (2020)]{Bice20} Bice, C.P. \& Toomre, J. 2020, 
\href{https://doi.org/10.3847/1538-4357/ab8190}{\apj},
\MYhref{https://ui.adsabs.harvard.edu/abs/2020ApJ...893..107B/abstract}{893, 107}

\bibitem[Brown et al. (2010)]{Brown10} Brown, B.P., Browning, M.K., Brun, A.S., Miesch, M.S., \& Toomre, J. 2010,
\href{http://dx.doi.org/10.1088/0004-637X/711/1/424}{\apj}, 
\MYhref{https://ui.adsabs.harvard.edu/abs/2010ApJ...711..424B/abstract}{711, 424}

\bibitem[Brun \& Toomre (2002)]{Brun02} Brun, A.S. \& Toomre, J., 2002, 
\href{https://doi.org/10.1086/339228}{\apj},
\MYhref[blue]{http://adsabs.harvard.edu/abs/2002ApJ...570..865B}{570, 865}

\bibitem[Brun et al. (2010)]{Brun10} Brun, A.S., Antia, H.M., Chitre, S.M. 2010, 
\href{http://doi.org/10.1051/0004-6361/200913166}{\apj},
\MYhref[blue]{https://ui.adsabs.harvard.edu/abs/2010A\&A...510A..33B/abstract}{510, A33}

\bibitem[Brun et al. (2011)]{Brun11} Brun, A.S., Miesch, M.S., \& Toomre, J. 2002, 
\href{https://doi.org/10.1088/0004-637X/742/2/79}{\apj},
\MYhref[blue]{https://ui.adsabs.harvard.edu/abs/2011ApJ...742...79B/abstract}{742, 79}

\bibitem[Busse (2002)]{Busse02} Busse, F.H., 2002, 
\href{https://doi.org/10.1063/1.1455626}{Phys. Fluids}, \MYhref[blue]{http://adsabs.harvard.edu/abs/2002PhFl...14.1301B}{14, 4}

\bibitem[Chandrasekhar (1961)]{Chandrasekhar61} Chandrasekhar, S. 1961, Hydrodynamic and Hydromagnetic Stability (Oxford: Clarendon)



\bibitem[Christensen-Dalsgaard et al. (1996)]{Dalsgaard96} Christensen-Dalsgaard, J., Dappen, W., Ajukov, S.V. et al. 1996,
\href{https://doi.org/10.1126/science.272.5266.1286}{Sci}, \MYhref{http://adsabs.harvard.edu/abs/1996Sci...272.1286C}{272, 1286}

\bibitem[Collins (1963)]{Collins63} Collins, G.W. II 1963, Astrophysical Journal, 
\href{https://doi.org/10.1086/147712}{\apj},
\MYhref{https://ui.adsabs.harvard.edu/abs/1963ApJ...138.1134C/abstract}{138, 1134}
 
\bibitem[Edwards (1990)]{Edwards90} Edwards, J.M. 1990,
\href{https://doi.org/10.1080/03091929008208942}{GAFD},
\MYhref{https://ui.adsabs.harvard.edu/abs/1990GApFD..55....1E/abstract}{55, 1}

\bibitem[Elliot et al. (2000)]{Elliot00} Elliot, J.R., Miesch, M.S. \& Toomre, J. 2000, 
\href{https://doi.org/10.1086/308643}{\apj}, 
\MYhref{http://adsabs.harvard.edu/abs/2000ApJ...533..546E}{533, 546}


\bibitem[Featherstone \& Hindman (2016a)]{Featherstone16a} Featherstone, N.A. \& Hindman, B.W. 2016a,  
\href{https://doi.org/10.3847/0004-637X/818/1/32}{\apj}, 
\MYhref[blue]{http://adsabs.harvard.edu/abs/2016ApJ...818...32F}{818 (1), 38}

\bibitem[Featherstone \& Hindman (2016b)]{Featherstone16b} Featherstone, N.A. \& Hindman, B.W. 2016b, 
\href{https://doi.org/10.3847/2041-8205/830/1/L15}{\apjlett}, 
\MYhref{http://adsabs.harvard.edu/abs/2016ApJ...830L..15F}{830, L15}

\bibitem[Featherstone (2018)]{Featherstone18} Featherstone, N. 2018, Rayleigh 0.9.1, doi: \href{http://doi.org/10.5281/zenodo.1236565}{http://doi.org/10.5281/zenodo.1236565}
\bibitem[Gastine et al. (2013)]{Gastine13} Gastine, T., Wicht, J. \& Aurnou, J.M. 2013, 
\href{https://doi.org/10.1016/j.icarus.2013.02.031}{\icarus}, 
\MYhref[blue]{http://adsabs.harvard.edu/abs/2013Icar..225..156G}{225, 156}



\bibitem[Gilman \& Glatzmaier (1981)]{Gilman81} Gilman, P.A. \& Glatzmaier, G.A. 1981, 
\href{https://doi.org/10.1086/190714}{\apjs},
\MYhref[blue]{http://adsabs.harvard.edu/abs/1981ApJS...45..335G}{45, 335}

\bibitem[Gilman (1983)]{Gilman83} Gilman, P.A. 1983, \apj,
\MYhref[blue]{https://ui.adsabs.harvard.edu/abs/1983ApJS...53..243G/abstract}{53, 243}


\bibitem[Greer (2015)]{Greer15} Greer, B.J., Hindman, B.W., Featherstone, N.A. \& Toomre, J. 2015,
\href{https://doi.org/10.3847/0004-637X/824/2/128}{\apjl},
\MYhref{http://adsabs.harvard.edu/abs/2016ApJ...824..128G}{803, L17}

\bibitem[Gough (1969)]{Gough69} Gough, D.O. 1969, \href{https://doi.org/10.1175/1520-0469(1969)026<0448:TAAFTC>2.0.CO;2}{J. Atmos. Sci.}, 
\MYhref[blue]{http://adsabs.harvard.edu/abs/1969JAtS...26..448G}{26, 448}

\bibitem[Guerrero et al. (2016)]{Guerrero16} Guerrero, G., Smolarkiewicz, E.W., de Gouveia Dal Pino, E.M., Kosovichev, A.G., \& Mansour, N.N. 2016, \href{https://doi.org/10.3847/2041-8205/828/1/L3}{\apjl}, 
\MYhref[blue]{https://ui.adsabs.harvard.edu/abs/2016ApJ...828L...3G/abstract}{828, L3}

\bibitem[Hanasoge et al. (2012)]{Hanasoge12} Hanasoge, S.M., Duvall, T.L., Jr., \& Sreenivasan, K.R. 2012, 
PNAS,
109, 11928

\bibitem[Hindman et al. (2020)]{Hindman20} Hindman, B.W., Featherstone, N.A., \& Julien, K. 2020, \apj, submitted

\bibitem[Hotta et al. (2015)]{Hotta15} Hotta, H., Rempel, M. \& Yokoyama, T. 2015, 
\href{https://doi.org/10.1088/0004-637X/798/1/51}{\apj}, 
\MYhref{https://adsabs.harvard.edu/abs/2015ApJ...798...51H}{798, 51}

\bibitem[Hotta et al. (2019)]{Hotta19} Hotta, H., Iijimi, H., \& Kusano, K. 2019, 
\href{https://doi.org/10.1126/sciadv.aau2307}{Sci Adv}, 
\MYhref{https://ui.adsabs.harvard.edu/abs/2019SciA....5.2307H/abstract}{5, eaau2307}

\bibitem[Howe et al. (2000)]{Howe00} Howe, R., Christensen-Dalsgaard, J., Hill, F., et al. 2000, 
\href{https://doi.org/10.1126/science.287.5462.2456}{Sci}, 
\MYhref[blue]{http://adsabs.harvard.edu/abs/2000Sci...287.2456H}{287, 2456}

\bibitem[Howe et al. (2005)]{Howe05} Howe, R., Christensen-Dalsgaard, J., Hill, F., et al. 2005, 
\href{https://doi.org/10.1086/497107} {\apj},
\MYhref{https://ui.adsabs.harvard.edu/abs/2005ApJ...634.1405H/abstract}{634, 1405}

\bibitem[Hurle et al. (1966)]{Hurle66} Hurle, D.T.J., Jakeman, E., \& Pike, E.R. 1966,
\href{https://doi.org/10.1098/rspa.1967.0039}{Proc. R. Soc. Lond. A}, 
\MYhref{https://ui.adsabs.harvard.edu/abs/1967RSPSA.296..469H/abstract}{296, 469}
	
\bibitem[Jones et al. (2009)]{Jones09} Jones, C.A., Kuzanyan, K.M., \& Mitchell, R.H. 2009,
\href{https://doi.org/10.1017/S0022112009007253}{J. Fl. Mech.},
\MYhref{https://ui.adsabs.harvard.edu/abs/2009JFM...634..291J/abstract}{634, 291}

\bibitem[Jones et al. (2011)]{Jones11} Jones, C.A., Boronski, P., Brun, A.S., et al. 2011, 
\href{https://doi.org/10.1016/j.icarus.2011.08.014}{\icarus},
\MYhref{http://adsabs.harvard.edu/abs/2011Icar..216..120J}{216, 120}
 
\bibitem[O'Mara et al. (2016)]{OMara16} O'Mara, B., Miesch M.S., Featherstone, N.A., \& Augustson, K.C. 2016,\
\href{https://dx.doi:10.1016/j.asr.2016.03.038}{AdSpR},
\MYhref{https://ui.adsabs.harvard.edu/abs/2016AdSpR..58.1475O/abstract}{58, 1475}


\bibitem[Matilsky et al. (2019)]{Matilsky19} Matilsky, L.I., Hindman, B.W., \& Toomre, J. 2019, 
\href{https://doi.org/10.3847/1538-4357/aaf647}{\apj}, 
\MYhref{https://ui.adsabs.harvard.edu/abs/2019ApJ...871..217M/abstract}{871, 217}

\bibitem[Matilsky \& Toomre (2020)]{Matilsky20} Matilsky, L.I. \& Toomre, J. 2020, 
\href{https://doi.org/10.3847/1538-4357/ab791c}{\apj},
\MYhref{https://ui.adsabs.harvard.edu/abs/2020ApJ...892..106M/abstract}{892, 106}

\bibitem[Matsui et al. (2016)]{Matsui16} Matsui, H., Heien, E., Aubert, J., Aurnou, J.M., Avery, M. et al., 2016,\
\href{https://doi.org/10.1002/2015GC006159}{Geochem. Geophys.}, 
\MYhref{http://adsabs.harvard.edu/abs/2016GGG....17.1586M}{17, 1586}

\bibitem[Miesch et al. (2006)]{Miesch06} Miesch, M.S. Brun, A.S., \& Toomre, J. 2006
\href{https://doi.org/10.1086/499621}{\apj}, 
\MYhref{https://ui.adsabs.harvard.edu/abs/2006ApJ...641..618M/abstract}{641, 618}


\bibitem[Nelson et al. (2018)]{Nelson18} Nelson, N.J., Featherstone, N.A., Miesch, M.S., \& Toomre, J. 2018
\href{https://doi.org/10.3847/1538-4357/aabc07}{\apj}, 
\MYhref{https://ui.adsabs.harvard.edu/abs/2018ApJ...859..117N/abstract}{859, 117}

\bibitem[Rast et al. (2008)]{Rast08} Rast, M.P., Ortiz, A., \& Meisner, R.W. 2008,
\href{https://doi.org/10.1086/524655}{\apj}, 
\MYhref{https://ui.adsabs.harvard.edu/abs/2008ApJ...673.1209R/abstract}{673, 1209}


\bibitem[Thompson et al. (2003)]{Thompson03} Thompson, M. J., Christensen-Dalsgaard, J., Miesch, M. S., \& Toomre, J. 2003,
\href{https://doi.org/10.1146/annurev.astro.41.011802.094848}{\araa},
\MYhref{https://ui.adsabs.harvard.edu/abs/2003ARA\%26A..41..599T/abstract}{41, 599}

\bibitem[Unno et al.  (1989)]{Unno89} Unno, W., Osaki, Y., Ando, H., et al. 1989, Nonradial Oscillations of Stars, 2nd ed. (University of Tokyo Press)

\bibitem[von Zeipel (1924)]{vonZeipel24} von Zeipel, H. 1924, 
\href{https://doi.org/10.1093/mnras/84.9.665}{\mnras},
\MYhref{https://ui.adsabs.harvard.edu/abs/1924MNRAS..84..665V/abstract}{84, 665}

\end{thebibliography}
\end{document}